\pdfoutput=1
\documentclass[aps,pra,pdftex,floatfix,intlimits,showkeys,showpacs,twocolumn,superscriptaddress,a4paper]{revtex4-1}

\usepackage{amsfonts,amsmath,amssymb}
\usepackage{dcolumn}
\usepackage[draft]{fixme}
\usepackage{graphicx}
\usepackage[utf8x]{inputenc}
\usepackage{textcomp}
\usepackage{hypernat}
\usepackage{hyperref}
\usepackage{breakurl}

\fxsetup{layout=pdfcnote,mode=multiuser}
\FXRegisterAuthor{ff}{aff}{FF}
\FXRegisterAuthor{jk}{ajk}{JK}
\FXRegisterAuthor{sp}{asp}{SP}

\begin{document}

\title{Rotational-state-specific guiding of large molecules}%

\author{Stephan Putzke}%
\author{Frank Filsinger}%
\affiliation{Fritz-Haber-Institut der Max-Planck-Gesellschaft, Faradayweg 4-6, 14195 Berlin, Germany}%
\author{Henrik Haak}%
\affiliation{Fritz-Haber-Institut der Max-Planck-Gesellschaft, Faradayweg 4-6, 14195 Berlin, Germany}%
\author{Jochen K\"upper}%
\affiliation{Fritz-Haber-Institut der Max-Planck-Gesellschaft, Faradayweg 4-6, 14195 Berlin, Germany}%
\affiliation{Center for Free-Electron Laser Science, DESY, Notkestrasse 85, 22607 Hamburg, Germany}%
\affiliation{University of Hamburg, Luruper Chaussee 149, 22761 Hamburg, Germany}%
\author{Gerard Meijer}%
\email[Electronic mail: ]{meijer@fhi-berlin.mpg.de}%
\affiliation{Fritz-Haber-Institut der Max-Planck-Gesellschaft, Faradayweg 4-6, 14195 Berlin, Germany}%

\date{\today}%

\begin{abstract}\noindent%
A beam of polar molecules can be focused and transported through an ac electric quadrupole guide. At
a given ac frequency, the transmission of the guide depends on the mass-to-dipole-moment (m/\textmu) 
ratio of the molecular quantum state. Here we present a detailed characterization of the 
m/\textmu~selector, using a pulsed beam of benzonitrile (C$_6$H$_5$CN) molecules in combination with 
rotational quantum state resolved detection. The arrival time distribution as well as the transverse velocity 
distribution of the molecules exiting the selector are measured as a function of ac frequency. The
\textmu/$\Delta$\textmu~resolution of the selector can be controlled by the applied ac waveforms and
a value of up to 20 can be obtained with the present setup. This is sufficient to exclusively transmit 
molecules in the absolute ground state of benzonitrile, or rather in quantum states that have the same 
m/\textmu~value as the ground state. The operation characteristics of the m/\textmu~selector are in 
quantitative agreement with the outcome of trajectory simulations. 
\end{abstract}%
\maketitle%

\section{Introduction}
\label{sec:intro}
Molecular beams have played central roles in many experiments in physics and chemistry and have found a wide 
range of applications~\cite{Scoles:MolBeam:1and2}. For many of these applications it is crucial that the internal quantum
state distribution of the molecules in the beam is accurately known and can be controlled. Cooling of the molecules
in an adiabatic expansion strongly reduces the number of populated levels, thereby tremendously simplifying the
optical spectra of the molecules, for instance~\cite{Levy:Science214:263}. For large molecules, however, many levels 
will still be populated at the low temperatures ($\approx$1~K) that can be achieved in a molecular beam. 
Moreover, even at the lowest possible temperatures, molecules with a complex potential energy landscape like 
bio-molecules, are known to be frozen into different minima and to be present in various conformational structures 
in the beam~\cite{Rizzo:JCP83:4819}. To control the quantum-state distribution (and thereby the conformational 
distribution) of large molecules in a beam, therefore, additional filtering techniques need to be applied. This can be 
achieved with electric, magnetic or optical fields, or any combination thereof~\cite{Meerakker:NatPhys4:595}. In the 
work presented here we discuss the quantum state selection of a beam of large molecules using electric fields.

A wide variety of electric field geometries has been used in the past to manipulate the trajectories of polar molecules
in a beam. The original geometries were devised to create strong electric field gradients on the beam axis to efficiently
deflect molecules. Later, electric field geometries were designed to focus molecules in selected quantum states further
downstream. For these experiments, one has to distinguish between molecules that are in so-called low-field-seeking
or high-field-seeking quantum states; as the name suggests, the former are attracted to regions of low electric field 
whereas the latter experience a force in the direction of a high electric field. In a seminal experiment, performed more than
half a century ago, an electrostatic quadrupole focuser was used to couple a beam of ammonia molecules into a microwave 
cavity. This beam initially contained an almost equal number of molecules in the upper (low-field-seeking) and the 
lower (high-field-seeking) level of the $\mid J,K \rangle = \mid1,1\rangle$ inversion doublet. The radially symmetric 
electric field in the quadrupole focuser had a zero of the electric field strength on the beam axis, resulting in a 
focusing (defocusing) of ammonia molecules in the upper (lower) inversion doublet level. The resulting inverted 
population distribution in the microwave cavity led to the demonstration of microwave amplification by stimulated
emission of radiation, i.e., the invention of the maser, by Gordon, Zeiger and Townes~\cite{Gordon:PR99:1264}.

Focusing is considerably more complicated for molecules in high-field-seeking states than for molecules in low-field-seeking 
states as no electric field maximum can be created on the molecular beam axis using static fields alone. Due to their small 
rotational constants and the resulting high density of levels, however, large molecules are high-field seeking in all their 
low-lying rotational levels already at modest electric field strengths. To focus large molecules, therefore, dynamic focusing has 
to be applied~\cite{Auerbach:JCP45:2160}, much like what is routinely done to transport charged 
particles~\cite{Courant:AnnPhys3:1}. Dynamic focusing of neutral polar molecules has first been experimentally demonstrated 
on ammonia molecules in high-field-seeking quantum states~\cite{Kakati:PLA24:676}. More
recently, it has been applied in the deceleration of metastable CO~\cite{Bethlem:PRL88:133003} in the trapping of ND$_3$ 
molecules~\cite{Veldhoven:PRL94:083001}, in the guiding of a cold beam of molecules from an effusive source~\cite{Junglen:PRL92:223001} 
and in the transport of CaF~\cite{Wall:PRA80:043407}, for instance. Dynamic focusing has also been applied to large molecules, e.g.,  
in the deceleration of benzonitrile~\cite{Wohlfart:PRA77:031404} as well as in the
separation of two different conformational structures of aminophenol~\cite{Filsinger:PRL100:133003}. In the latter experiment
the ac electric quadrupole guide was used to selectively transmit molecules depending on their mass-to-dipole-moment 
(m/\textmu) ratio, analogous to the transmission of ions with a certain mass-to-charge (m/q) ratio in a quadrupole mass filter.

Here we present a detailed characterization of a second generation m/\textmu~selector.  A pulsed beam of 
rotationally cold benzonitrile (C$_6$H$_5$CN) molecules is injected into the ac electric quadrupole selector, 
and the arrival time distribution as well as the transverse velocity distribution of the molecules exiting the selector 
are measured using high-resolution electronic spectroscopy. The dependence of the \textmu/$\Delta$\textmu{} 
resolution of the selector on the applied ac waveforms is studied, and the transmission of selected subsets of 
rotational levels is demonstrated. The operation characteristics of the m/\textmu~selector are compared to 
the outcome of numerical trajectory simulations. 

\section{Experimental Setup}
\label{sec:setup}

\begin{figure}
   \centering
   \includegraphics[width=\linewidth]{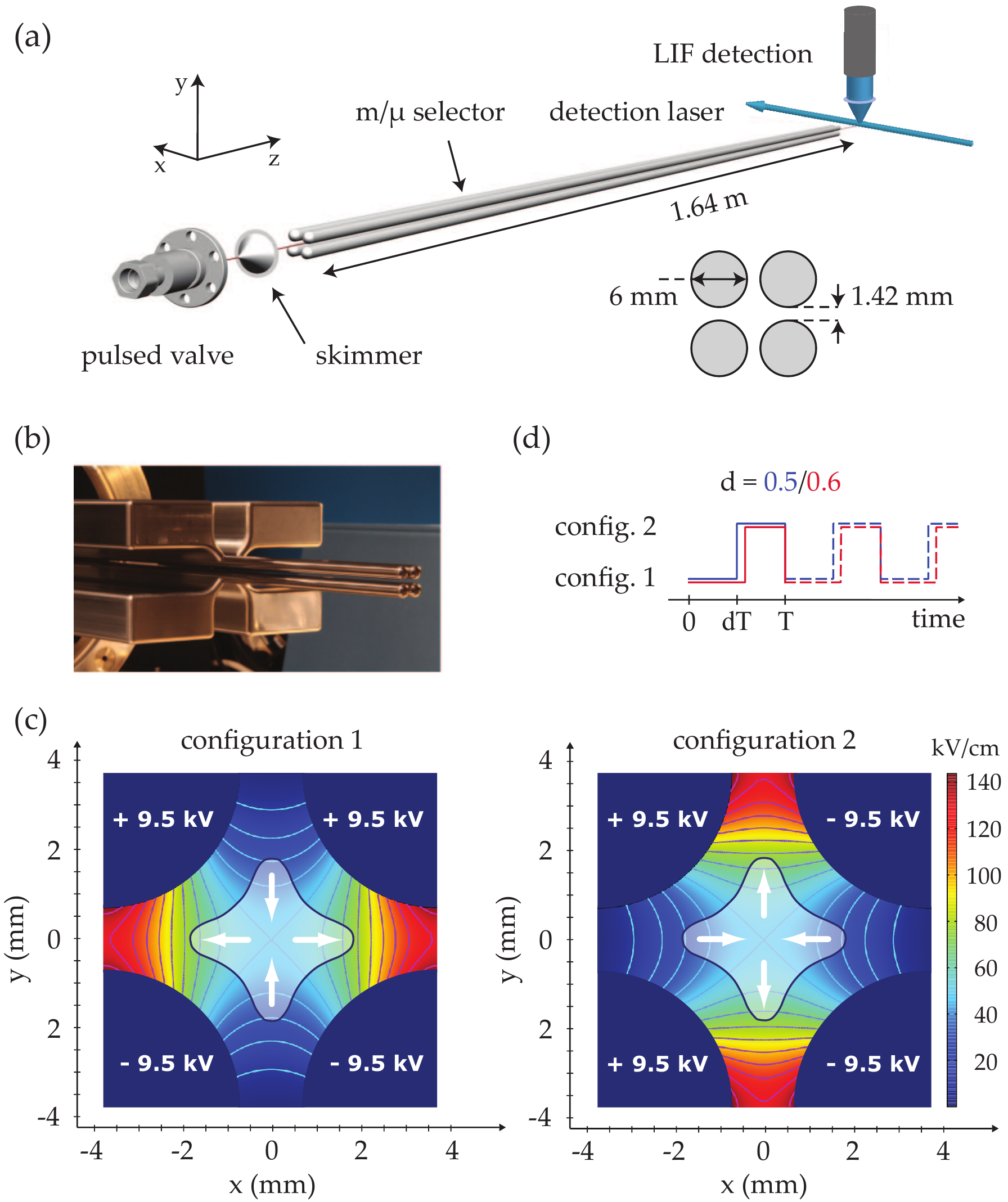}
   \caption{(a) Schematic view of the experimental setup. (b) Photograph of the end of the
      selector, showing how the cylindrical electrodes are mounted to the four rigid backbones. 
      (c) Electric field for either one of the two high-voltage configurations. The electric field at
      the center is 48~kV/cm and contour lines are drawn every 10~kV/cm away from the center. The transparent 
      cross-shaped overlay shows the area in the (x,y)-plane to which the molecules are confined. 
      (d) Switching scheme for two different duty cycles at the same ac frequency ($1/T$).}
   \label{fig:results:setup}
\end{figure}

A scheme of the experimental setup is shown in \autoref{fig:results:setup}(a). Argon gas at 2.5~bar is bubbled through 
liquid benzonitrile in a room-temperature container. The resulting gas mixture of 0.04\% benzonitrile in Argon is expanded 
through a 0.8~mm diameter nozzle into vacuum using a pulsed valve (General Valve, Series 99) at a repetition rate 
of 40~Hz. A beam of benzonitrile molecules with a rotational temperature of about 1~K and with a mean forward velocity 
of around 570~m/s is thus produced. The value for the most probably velocity is not precisely known but is expected to 
be close to, but probably slightly higher than, that for a mono-atomic ideal gas expanding from a room-temperature 
valve~\cite{Scoles:MolBeam:1and2}; with $v_p$ = $\sqrt{5 k_B T_0/m}$ with $k_B$ the Boltzmann constant, $m$ the 
mass of the Ar atoms and $T_0$=295~K, a lower limit of $v_p$ = 552~m/s is obtained. For a detailed analysis of the 
experimental results, the initial length of the gas pulse and the width of the velocity distribution are other important 
parameters. We do not know these independently, but we can accurately determine the relative spread in the arrival 
time distribution at the end of the beam machine as being 4.8\% ($\it{vide}$ $\it{infra}$). We mainly attribute this to the 
width of the velocity distribution, which is assumed to have a full-width-at-half maximum (FWHM) of about 25~m/s, 
corresponding to a relative velocity spread of about 4.4\%. This narrow velocity distribution corresponds to a longitudinal 
translational temperature of 0.5--0.6~K.

After passing through a 1.5~mm diameter skimmer (Beam Dynamics, model 2) placed about 40~mm from the 
nozzle, the molecules enter the selector 32~mm behind the tip of the skimmer. The selector consists of four 1.64~m 
long and 6~mm diameter stainless steel electrodes. The ends of the highly polished electrodes are rounded off with a 
radius of curvature of 3~mm. The electrodes are centered at the corners of a square with a diagonal of 10.5~mm. This 
implies that the closest distance of the surface of each of the four electrodes to the molecular beam axis is 2.25~mm; 
the closest distance between adjacent electrodes is 1.42~mm, as indicated in \autoref{fig:results:setup}(b). To achieve 
the best possible performance of the selector care is taken to keep the distances between the electrodes as constant 
as possible over the full length of the device. To this end, four rigid 1.4~m long aluminum holders have been machined 
that serve as straight, stable backbones on which the electrodes are mounted. The electrodes, which are out of
magnetic stainless steel, are attached via small magnets that are inserted from the backside into the holders. This 
mounting scheme prevents the unavoidable bending of the electrodes that would occur when screw holes would be 
drilled into them and it also prevents mechanical stress that would be present when the electrodes would be screwed 
onto the holders. In the completely mounted setup, we can only accurately measure the spacing between adjacent 
electrodes. It is found that this spacing deviates less than $\pm$0.02~mm from its nominal value of 1.42~mm over the 
full length of the selector.

High voltages are connected to the four electrodes of the selector according to either one of the configurations shown in 
\autoref{fig:results:setup}(c). In configuration 1, voltages of +9.5~kV are applied to the upper two electrodes while the 
lower two electrodes are at -9.5~kV, creating a saddle-point of the electric field of 48~kV/cm on the molecular beam axis. 
In the horizontal plane, the electric field increases with distance from the center whereas it decreases away from the center
in the vertical plane. Benzonitrile molecules in high-field-seeking states, therefore, will be focused in the vertical direction 
while being defocused horizontally, as indicated by the white arrows in \autoref{fig:results:setup}(c). In going from 
configuration 1 to configuration 2, the voltage applied to the right upper (left lower) electrode is switched from +9.5~kV 
to -9.5~kV (-9.5~kV to +9.5~kV), effectively rotating the field configuration, and thereby the (de-)focusing direction, by 90 
degrees. By switching back and forth between these two configurations, benzonitrile molecules can be transported on
stable trajectories through the selector. The switching between the configurations is performed with two push-pull 
high-voltage switches (Behlke, HTS 301-03-GSM) according to the scheme depicted in \autoref{fig:results:setup}(d). The
time-dependence is shown there as a block-function; in reality it takes about 0.5~\textmu s to reach the final 
voltage. The ac switching frequency ($1/T$) is in the 2--8~kHz range and the fraction of the period during which 
configuration 1 is applied is denoted as the duty cycle $d$. By varying the duty cycle the \textmu/$\Delta$\textmu{}
resolution is changed~\cite{PhysRevA.82.052513} and in the present experiments values of $d$ in the range 0.5--0.6 have 
been used. The cross-shaped overlay in \autoref{fig:results:setup}(c) depicts the typical area in the (x,y)-plane to which 
molecules stay confined while flying in the z-direction. From this, benzonitrile molecules transmitted on stable trajectories 
are seen to sample electric field strengths in the range of 25--77~kV/cm.

The benzonitrile molecules are monitored 40~mm behind the end of the selector via laser induced fluorescence
(LIF) detection. Individual rotational transitions in the origin band of the electronic \mbox{$S_1\leftarrow S_0$} system are induced
using narrowband radiation from a continuous wave ring dye laser (Coherent 899-21) that is frequency doubled 
in an external cavity (Spectra Physics, LAS WaveTrain). Typically 15~mW of tunable 274~nm radiation with an instantaneous
bandwidth of less than 1~MHz is obtained in this way. The laser frequency is actively stabilized against a frequency 
stabilized Helium-Neon laser (SIOS SL 03), thus allowing to keep the frequency constant to within 2~MHz for several hours. 
The collimated 2.5~mm diameter laser beam intersects the molecular beam under a right angle in the horizontal plane while the 
fluorescence is detected from above, i.e., along the y-direction. The transmission of benzonitrile molecules through 
the selector in individual rotational $J_{K_aK_c}$ levels is measured via time-resolved counting of the LIF photons.
In this setup, we are also sensitive to the transverse velocity distribution of the benzonitrile molecules in the 
x-direction via the Doppler-profile of the rotational transitions.

\section{Experimental results}
\label{sec:results}

\begin{figure}
   \centering
   \includegraphics[width=\linewidth]{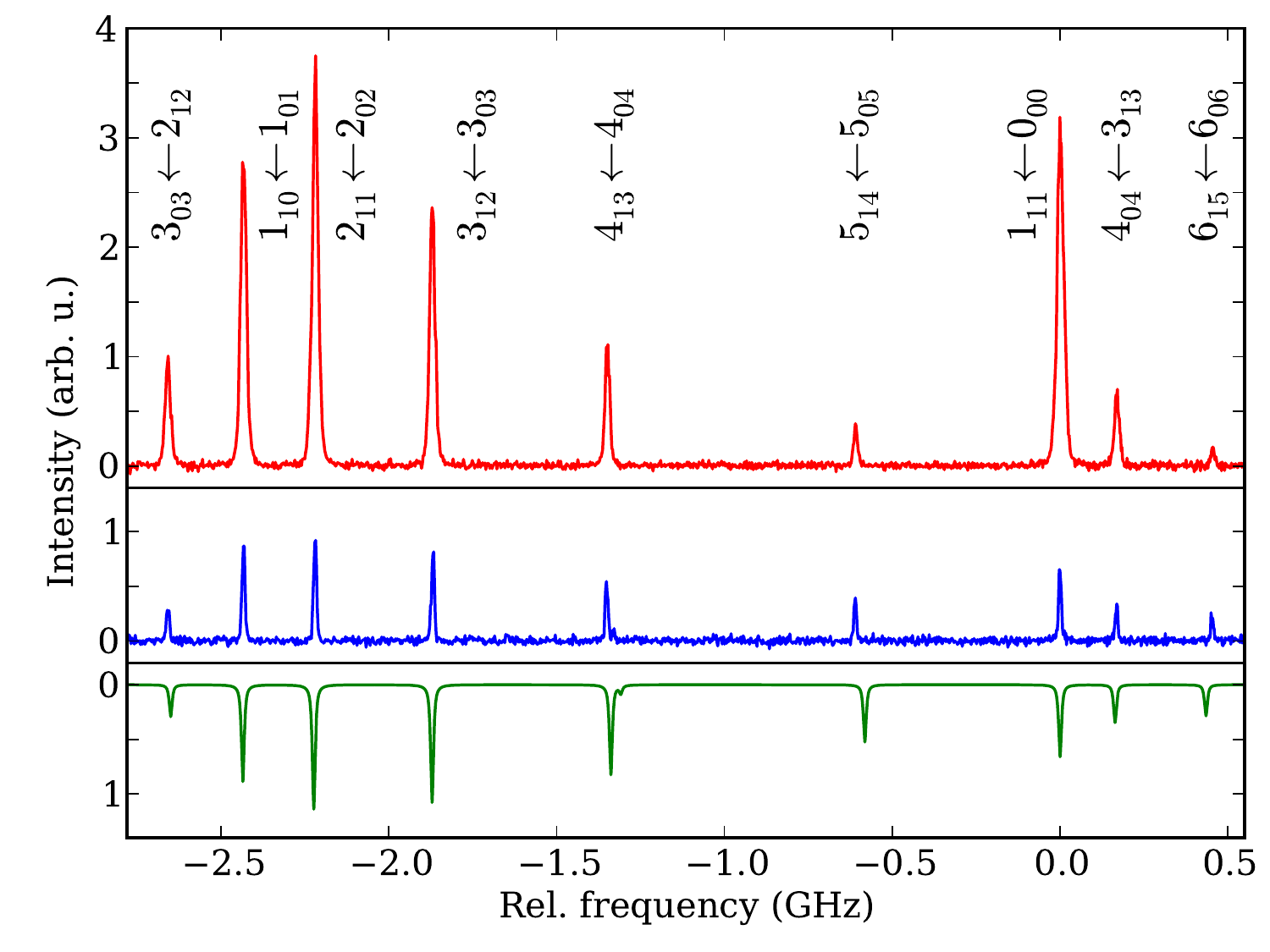}
   \caption{Rotationally resolved fluorescence excitation spectra of benzonitrile. The horizontal axis is relative to the
      frequency of the $1_{11}\leftarrow0_{00}$ transition at 36512.74~cm$^{-1}$. Middle trace: Spectrum obtained 
      without voltages applied to the selector, i.e., in free flight. Lower trace: Simulated free-flight spectrum (plotted upside 
      down), assuming a rotational temperature of 1.2~K. The intensities are scaled to the intensity of the $1_{11}\leftarrow0_{00}$ 
      transition in the middle trace and a Voigt profile with a Lorentzian contribution (FWHM) of 8~MHz and a Gaussian contribution
      (FWHM) of 3.7~MHz is taken. Upper trace: Spectrum obtained with the selector operating at an ac switching frequency of 
      3.8~kHz, $d$=0.50. The $J'_{K'_aK'_c} \leftarrow J_{K_aK_c}$ assignment for each of the rotational lines is indicated.}
  \label{fig:results:bnspectrafree}
\end{figure}

In \autoref{fig:results:bnspectrafree} a selected part of the rotationally resolved fluorescence excitation spectrum of
benzonitrile is shown. The fluorescence signal is integrated over a time-interval of 0.5~ms, thereby accumulating the 
signal from all benzonitrile molecules in a pulse. The spectrum shown in the middle trace is recorded in free flight, 
i.e., without voltages applied to the selector. A total of nine clearly separated rotational lines, originating from nine distinct 
low-lying rotational levels, 
are recognized. From this spectrum, the rotational temperature of benzonitrile in the molecular beam can be accurately
determined. Taking the known molecular constants of benzonitrile in the ground state~\cite{Wohlfart:JMolSpec247:119} 
and in the electronically excited state~\cite{Borst:CPL350:485} in combination with the degeneracy of the levels (spin statistics) and the 
rotational line intensities in the electronic transition~\cite{Western:pgopher}, the observed spectrum can be simulated 
with the rotational temperature as the only adjustable parameter. A simulated spectrum for a rotational temperature of 
1.2~K is shown in the lower trace, and is seen to match the experimental spectrum rather well. A Voigt profile with a Lorentzian
(Gaussian) contribution with a width (FWHM) of 8~MHz (3.7~MHz) is taken in the simulation. The width of the Lorentzian 
contribution has been taken from earlier high-resolution spectroscopy experiments~\cite{Borst:CPL350:485} and is consistent 
with our present observations. It should be noted that the radiative lifetime of the vibrationless level in the $S_1$ 
state is about 70~ns~\cite{Kobayashi:JCP86:1111, Lahmani:LaserChem10:41}, thus contributing less than 2.5~MHz to the observed
homogeneous linewidth; the remaining contribution has to result from pure dephasing, with an associated pure dephasing
time of about 60~ns. In the upper trace, the spectrum obtained with the ac quadrupole
selector operating at a switching frequency of 3.8~kHz, and with a duty cycle of 0.50, is shown. The fluorescence intensity 
is plotted on the same scale as for the spectrum recorded in free flight and an increase in peak intensity of almost a factor 
five is obtained for benzonitrile molecules in their rotational ground state ($0_{00}$) level. For rotational transitions starting 
from rotationally excited levels the intensity increase upon using the selector is seen to be smaller, however, and the intensity 
even decreases for the transitions starting from the $5_{05}$ and the $6_{06}$ levels.

\begin{figure}
   \centering
   \includegraphics[width=0.666\linewidth]{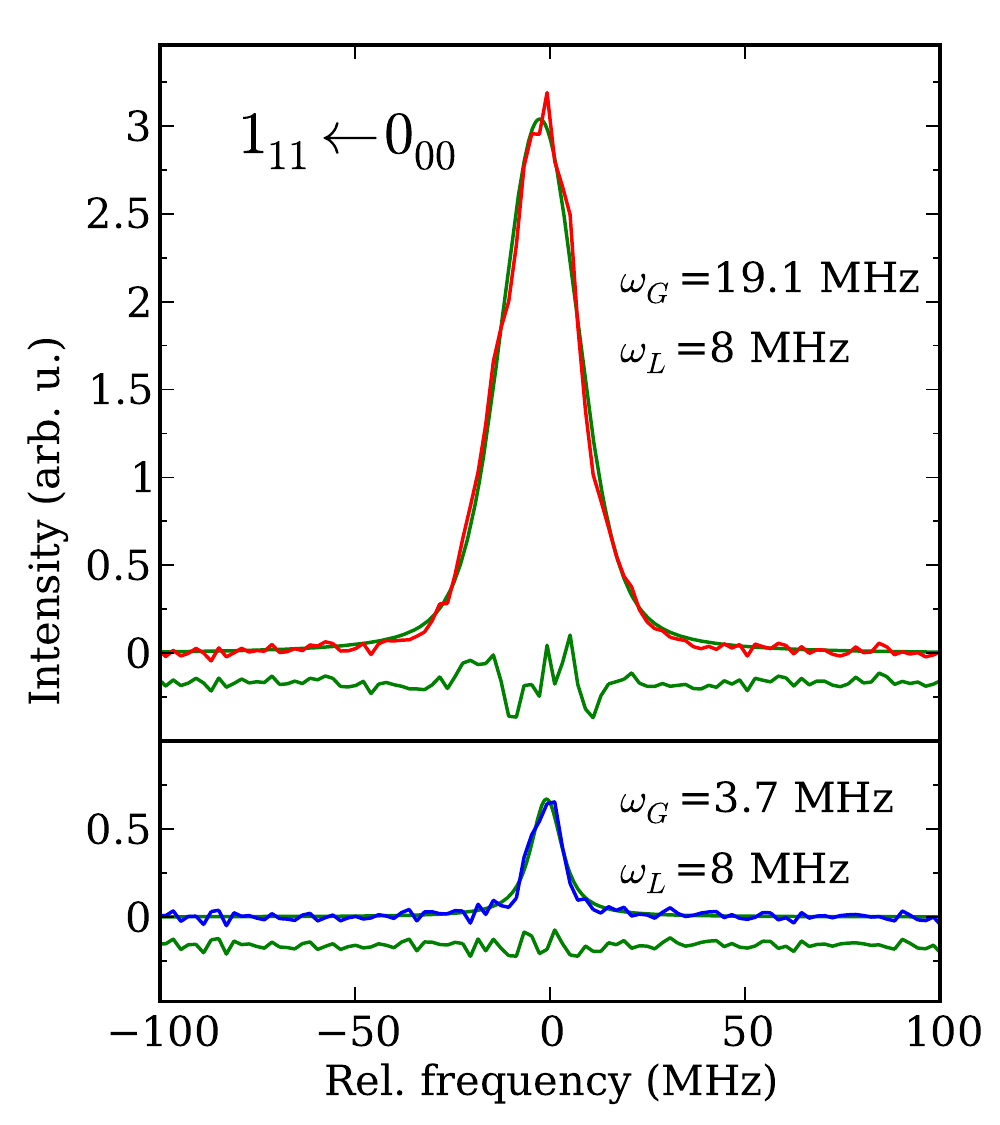}
    \caption{Expanded view of the $1_{11}\leftarrow0_{00}$ transition shown in \autoref{fig:results:bnspectrafree},
     measured with the selector on (upper trace) and off (lower trace). Voigt profiles fitted to the experimental 
     data are shown and the widths (FWHM) of the Lorentzian and Gaussian contributions to the Voigt profiles are 
     given. The difference of the observed and fitted profiles is shown underneath, slightly offset for clarity.}
  \label{fig:results:bnspectravoigt}
\end{figure}

It is evident from the spectra shown in \autoref{fig:results:bnspectrafree} that the rotational transitions are broadened when
the selector is switched on. This is seen more clearly in \autoref{fig:results:bnspectravoigt} where an expanded view of the 
$1_{11}\leftarrow0_{00}$ transition, recorded with the selector on (upper trace) and off (lower trace), is shown together with
the best fitting Voigt profiles. As the Lorentzian contribution to the profiles is kept fixed to 8~MHz, the width of the Gaussian 
contribution and the peak intensity are the only fitting parameters. A Gaussian contribution with a width of 3.7~MHz is 
obtained under free-flight conditions, determined by the residual Doppler broadening and the bandwidth of the laser system. 
When the selector is on, the width of the Gaussian contribution increases to 19.1~MHz due to the increased transverse 
velocity of the benzonitrile molecules exiting the selector. The spectrally integrated line intensity increases by a factor of 9.3 
when the selector is switched on. As the transverse spatial distribution of the molecules exiting the selector will be larger, 
the overlap with the detection laser will be somewhat smaller, implying that this factor 9.3 is a lower limit for the overall increase 
of the transmission of benzonitrile molecules in the $0_{00}$ level upon switching on the selector. 

\begin{figure}
   \centering
   \includegraphics[width=\linewidth]{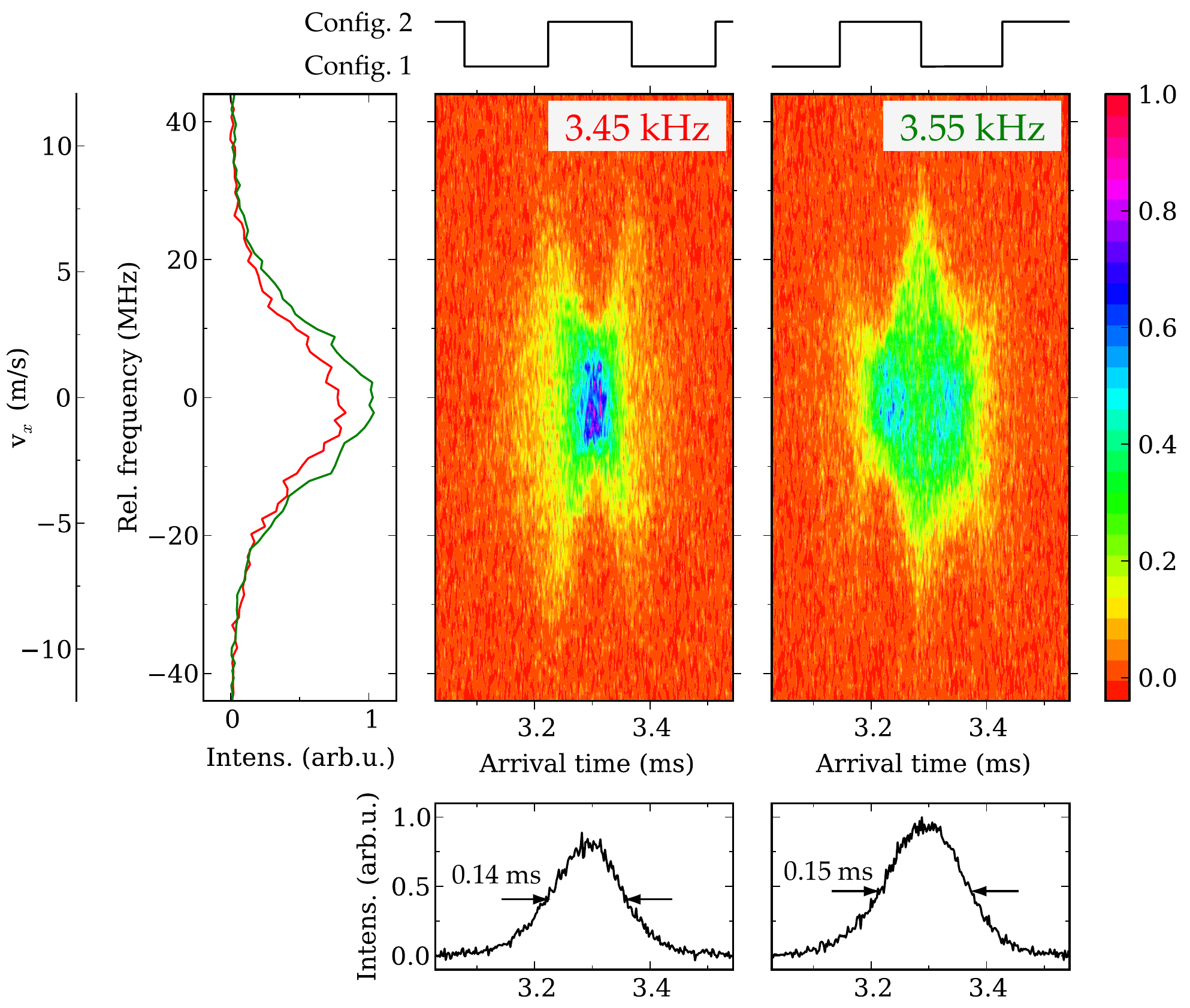}
   \caption{Intensity of the LIF signal, plotted in a false-colour representation, as a function of the frequency of the excitation 
     laser and as a function of arrival time for two different ac switching frequencies. The vertical axis is also 
     given in terms of the velocity in the x-direction. Above the contour-plots, the phase in the switching cycle at which the molecules 
     exited the selector is indicated. The panel on the left shows the time-integrated LIF signals as a function of $v_x$. The panels
     at the bottom show the arrival time distributions integrated over the Doppler-broadened profiles.}
  \label{fig:results:linescans}
\end{figure}

The spectral line-profiles shown in \autoref{fig:results:bnspectravoigt} are recorded by accumulating the fluorescence signal from
all ground-state benzonitrile molecules in the pulse. When these line profiles are detected for narrow time intervals, the 
time dependence of the transverse velocity distribution of the molecules exiting the selector can be measured. Such measurements
are shown in two-dimensional contour-plots in \autoref{fig:results:linescans}. For this, the intensity of the fluorescence 
signal, binned in 2~\textmu s time-intervals, is plotted in a false-colour representation as a function of the frequency of the 
excitation laser, which is scanned in 1~MHz steps from -40~MHz to +40~MHz around the center frequency of the 
$1_{11}\leftarrow0_{00}$ transition. Interesting structures are observed in these two-dimensional plots, that are clearly 
different when the selector is operated at 3.45~kHz (left contour plot) or at 3.55~kHz (right contour plot). The panel on the left 
shows the time-integrated LIF signals as a function of laser frequency. The Doppler shift on the vertical 
axis can be directly translated into a velocity component along the propagation direction of the laser, i.e., in the x-direction, and 
the corresponding velocity scale is given as well. The time on the horizontal axis is the time (in ms) since the triggering of the 
pulsed valve; the actual opening of the pulsed valve has a delay with respect to the trigger of approximately 0.2~ms. The panels
at the bottom show the arrival time distributions integrated over the Doppler-broadened profiles that are seen to have a FWHM
width of about 0.15~ms. 

In the experiments the switching sequence is started prior to the molecules entering the selector. The transverse velocity distribution 
that the molecules have in the detection region is mainly determined by the forces that the molecules experience near the end of 
the selector. For this reason, the phase in the switching cycle at which the molecules exit the selector, referred to hereafter as the 
end phase, is schematically shown above the contour-plots. As the detection of the molecules is performed only 40~mm from 
the end of the selector and as the width of the longitudinal velocity distribution is sufficiently narrow, the arrival time of the molecules 
in the detection region can be directly mapped to an end phase. 

In the measurements recorded with an ac switching frequency of 3.45~kHz, the molecules arriving at 3.3~ms exited the selector 
in the middle of configuration 2, i.e., during the middle of a horizontal focusing phase. This is actually at a turning point of the micro motion 
in the selector, and thus results in a rather parallel beam and a correspondingly narrow spectral line-profile. In the measurements 
recorded at the slightly higher frequency of 3.55~kHz, the molecules arriving at 3.3~ms exited the selector at the end of a horizontal 
focusing phase. These molecules therefore have the maximal velocity components along the x-direction, leading to the broadest 
line-profile; benzonitrile molecules exiting the selector at the middle of either configuration 2 or configuration 1 again have a narrow 
spectral profile. Related experiments to visualize the motion of dynamically confined neutral particles have been performed on 
metastable CO molecules passing through an alternating gradient decelerator~\cite{Bethlem:JPB39:R263} as well as on Rb atoms 
in an ac electric trap~\cite{Schlunk:PRL98:223002}.

\begin{figure}
   \centering
   \includegraphics[width=\linewidth]{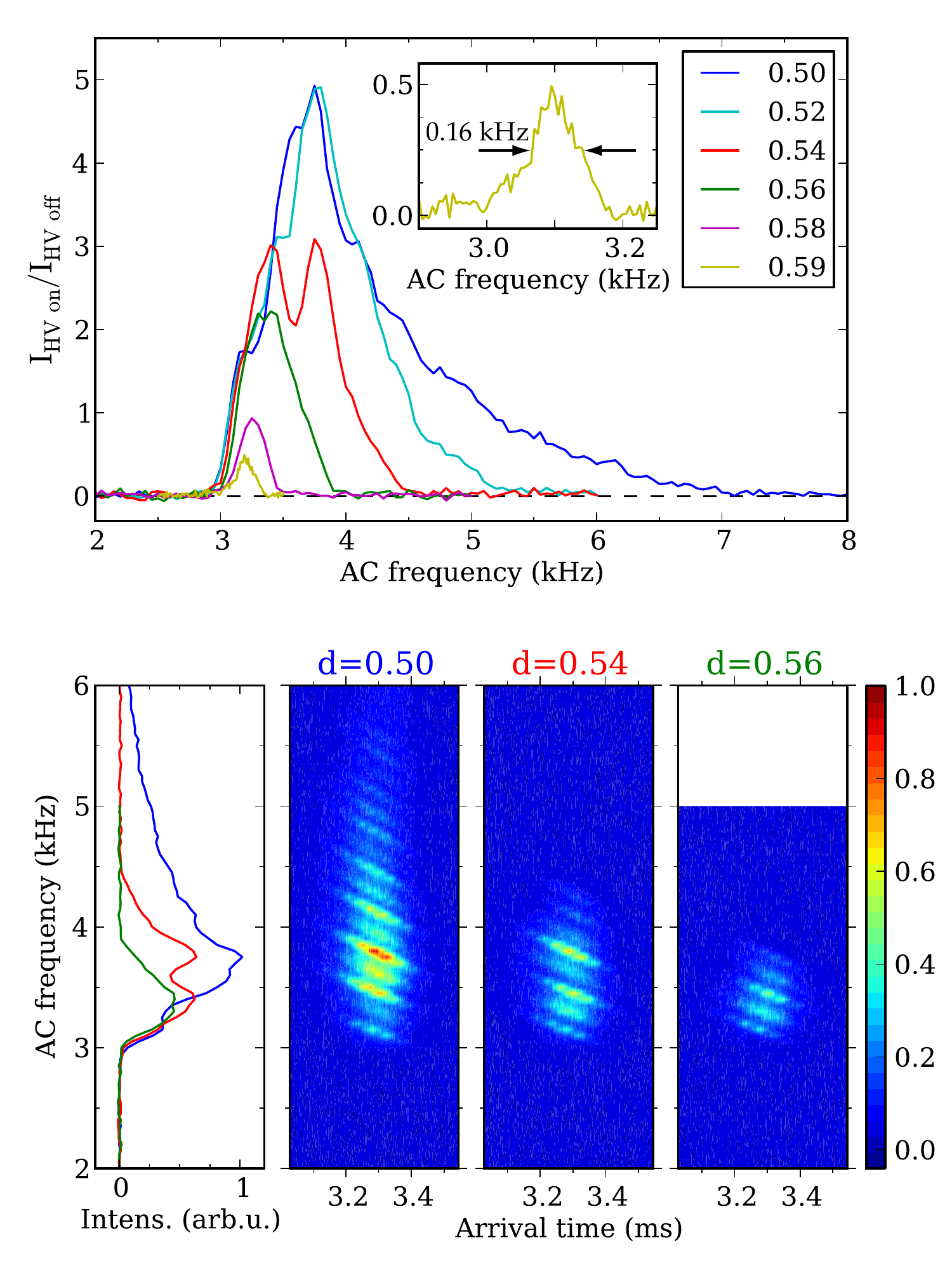}
   \caption{Upper panel: Time-integrated LIF signal intensity as a function of the ac switching frequency. Transmission curves are 
     measured at six different duty cycles, indicated in the inset, while the laser frequency is kept fixed to the center of the 
     $1_{11}\leftarrow 0_{00}$ transition. The transmission curve for $d$=0.59 is also shown on an expanded scale. The signal is 
     normalized to the LIF intensity obtained in free flight.      
     Lower panel: Measured intensity of the LIF signal, plotted in a false-colour representation, as a function of the ac switching 
     frequency and as a function of arrival time for three different duty cycles. The panel on the left shows the corresponding 
     time-integrated LIF signals as a function of ac switching frequency.}
  \label{fig:results:transmissioncurves}
\end{figure}

In the upper panel of \autoref{fig:results:transmissioncurves}, the time-integrated LIF signal of benzonitrile molecules in the $0_{00}$ 
level is shown as a function of the ac switching frequency in the 2--8~kHz range. The ac frequency is scanned
in steps of 50~Hz for all curves except for the one at a duty cycle of 0.59, which is recorded in 10~Hz steps; the latter curve is shown
once more on an expanded scale in the inset. The excitation laser is 
kept fixed to the center of the $1_{11}\leftarrow 0_{00}$ transition and the signal intensity is 
normalized to the signal that is obtained when no high voltages are applied to the selector. 
To have well-defined conditions in the experiment, the timing of the ac switching sequence was arranged such that 0.2~ms after the
trigger to the pulsed valve, i.e., at the moment that the pulsed valve opens, there is a fixed phase for all ac frequencies.
Although this implies that the phase that a benzonitrile molecule experiences at the entrance of the selector, the so-called 
start phase, is different for different ac frequencies this start phase can be precisely calculated once the position of the molecule in the 
pulse and its velocity is known. The end phase is in this situation different for different ac switching frequencies as well, leading to the 
observed intensity modulation on the individual curves. All the different curves show no transmission of the selector below a certain cut-off 
frequency, followed by a steep onset of the transmission around 3~kHz, a maximum at frequencies between 3--4~kHz and a gradual 
decrease in transmission for higher frequencies. This general behavior has been described before~\cite{Veldhoven:PRL94:083001}, 
and is similar to a typical transmission curve in a quadrupole mass filter, for instance.
At low ac switching frequencies the trajectories of the molecules are unstable, whereas at high frequencies the effective
potential well in which the molecules are confined becomes very shallow. Only in an intermediate frequency range good transmission 
on stable trajectories is achieved. The transmission of the m/\textmu~selector can be strongly modified when an asymmetric 
switching cycle is applied, i.e., when values of $d \neq0.50$ are chosen~\cite{PhysRevA.82.052513}. Transmission curves for six 
different duty cycles are shown to illustrate this. For a duty cycle of 0.50 the FWHM of the transmission curve for benzonitrile 
molecules in their ground-state level is about 1~kHz. With increasing duty cycle, the width of the transmission curve gets 
considerably narrower while the transmission intensity goes down. At a duty cycle of $d =0.59$, the FWHM of the transmission 
curve has come down to 0.16~kHz, while the peak transmission of the selector is then reduced by about a factor ten. 

In the lower panel of \autoref{fig:results:transmissioncurves}, the intensity of the measured LIF signal is shown in a false-colour 
representation as a function of ac switching frequency and as a function of arrival time for three selected duty cycles. The 
transmission curves shown in the upper panel have been obtained by time-integrating these signals, and three of these are 
shown again in the lower panel on the left. To understand the two-dimensional plots, one has to realize that at a given ac frequency 
and with the excitation laser at the center of the rotational transition, thereby only probing molecules with small velocities in the 
x-direction, the most strongly peaked arrival time distribution is obtained for molecules exiting the selector in the middle of 
configuration 2. This is directly evident from inspecting the contour-plots shown in \autoref{fig:results:linescans}. As shown there, 
at an ac frequency of 3.45~kHz, these benzonitrile molecules produce a narrow arrival time distribution, 
peaked at 3.3~ms. With increasing ac frequency, the corresponding peak in the arrival time distribution shifts to earlier times, 
explaining the slopes of the intensity profiles observed in \autoref{fig:results:transmissioncurves}. When the molecules exit the 
selector during the middle of configuration 1, i.e., during a vertical focusing phase, the overall intensity of the LIF signal is less. 
This can be understood from simple arguments: at the middle of either focusing phase, the beam coming out of the selector will 
be most parallel and its spatial distribution at the end of the selector will be maintained in the detection region. In the middle of a 
vertical focusing phase (configuration 1) the cloud of molecules is necessarily larger in the y-direction than in the middle of a 
vertical defocusing phase (configuration 2); this is actually the underlying operation principle of alternating gradient focusing. 
The spatial overlap with the laser beam will thus be smaller in the middle of configuration 1. Moreover, the velocity spread in 
the horizontal direction is larger in the middle of configuration 1 than in the middle of configuration 2. Both the decreased overlap 
in the vertical direction and the increased velocity spread in the horizontal one lead to a lower LIF intensity in the middle of 
configuration 1 than in the middle of configuration 2. This explains the occurrence of maxima with alternating intensity with 
increasing ac frequency, for a given arrival time. Of course, the profile of the arrival time distribution at any ac frequency is 
weighed by the intensity profile of the pulse, which is peaked around 3.3~ms and has a (FWHM) width of 0.15~ms. It is clear 
from these two-dimensional plots that when these are integrated over time to obtain the transmission curves as function of the 
ac frequency, some remnants of the structure originating from the varying end phases remain.

\begin{figure}
   \centering
   \includegraphics[width=\linewidth]{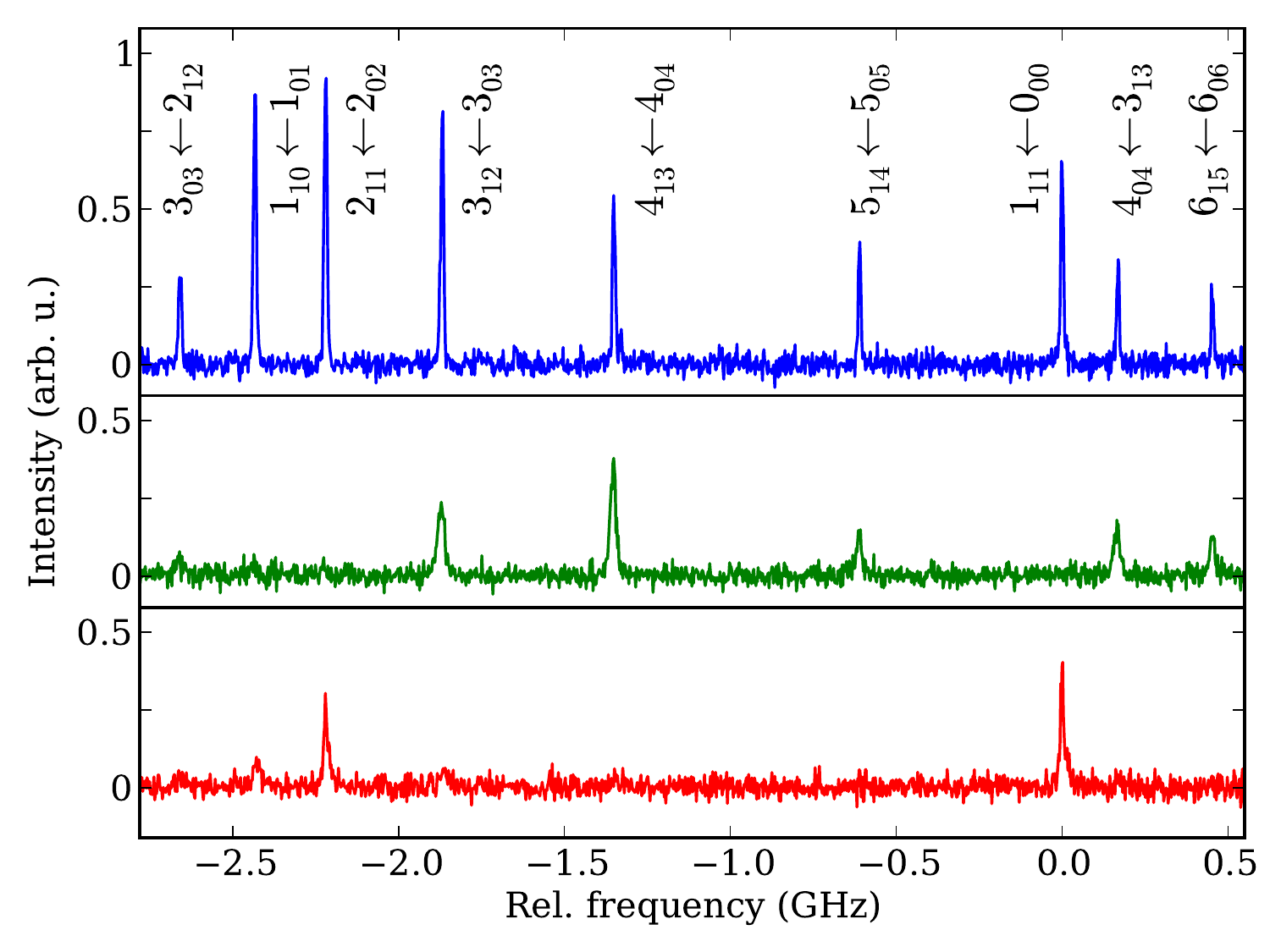}
   \caption{Rotationally resolved fluorescence excitation spectra of benzonitrile. Upper trace: Free-flight spectrum
      from \autoref{fig:results:bnspectrafree} in which the $J'_{K'_aK'_c} \leftarrow J_{K_aK_c}$ assignment for each 
      of the rotational lines is indicated. Middle trace: Spectrum obtained with the selector operating at an ac 
      switching frequency of 2.7~kHz, $d$=0.50. Lower trace: Spectrum obtained with the selector operating at an ac 
      switching frequency of 3.8~kHz, $d$=0.565.}
  \label{fig:results:bnspectraselection}
\end{figure}

The transmission curves shown in \autoref{fig:results:transmissioncurves} are for benzonitrile molecules in the $0_{00}$ 
ground-state level. The overall transmission profile shifts to lower ac switching frequencies when the forces on the molecules 
are weaker, i.e., for benzonitrile molecules in rotational levels with a lower dipole moment. This feature can be used to 
selectively transmit benzonitrile molecules in specific (subsets of) rotational quantum levels, as shown in 
\autoref{fig:results:bnspectraselection}. The upper spectrum shows once more the part of 
the rotationally resolved fluorescence excitation spectrum recorded without the use of the selector, i.e., the free-flight spectrum, 
shown already in \autoref{fig:results:bnspectrafree}. The spectrum in the middle trace demonstrates that molecules in the 
$0_{00}$, $1_{01}$ and $2_{02}$ rotational levels are not transmitted at all when the selector is operated at an ac switching 
frequency of 2.70~kHz and a duty cycle of 0.50. This switching frequency is below the abrupt low-frequency cut-off shown 
in \autoref{fig:results:transmissioncurves} and it is evident from the spectrum that only molecules in higher lying rotational
levels are transmitted under these conditions. To create a beam containing only ground-state benzonitrile molecules, the 
selector has to be operated at ac switching frequencies above those of the transmission maximum, under conditions that the 
transmission curve has a rather steep frequency cut-off on the high-frequency side as well. This has been done in the recording 
of the lower spectrum, during which the selector was operated at an ac switching frequency of 3.8~kHz and at a duty cycle of 0.565. 
In this case, only molecules in the $0_{00}$ and in the $2_{02}$ rotational levels as well as a very small fraction of molecules 
in the $1_{01}$ rotational level are transmitted. 

\section{Discussion}
\label{sec:simulations}

\begin{figure}
   \centering
   \includegraphics[width=\linewidth]{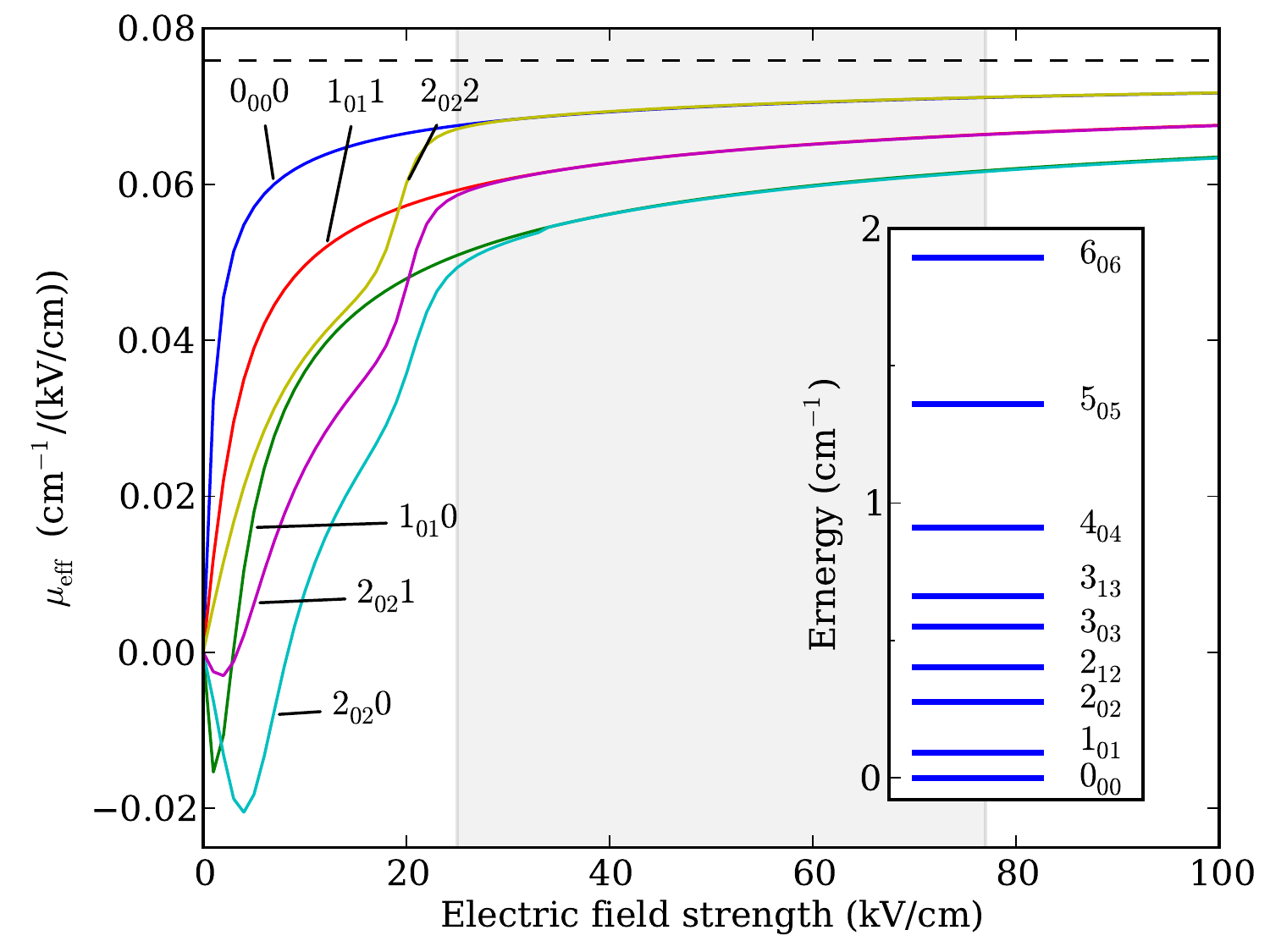}
   \caption{Effective dipole moments, \textmu$_{\textrm{eff}}$, of low rotational quantum states of benzonitrile, labeled as 
     $J_{K_aK_c}M$, as a function of the electric field strength. The dashed horizontal line indicates the 4.52~Debye permanent
     electric dipole moment of benzonitrile. The range of electric field strengths experienced by the 
     molecules while in the selector is indicated by the shaded area. In the inset the field-free energies of the nine rotational 
     levels probed in the measurements shown in \autoref{fig:results:bnspectrafree} are depicted.}
  \label{fig:results:efields}
\end{figure}

To quantitatively describe the operation characteristics of the selector we have to look in detail to the interaction of benzonitrile
in any of the low-lying rotational levels with electric fields. In the following, we will restrict ourselves to the $0_{00}$, the $1_{01}$ and
the $2_{02}$ levels, the lowest ones of the nine rotational levels probed in \autoref{fig:results:bnspectrafree}, whose field-free
energies are schematically depicted in the inset to \autoref{fig:results:efields}. In general, the force that neutral polar molecules 
experience in the electric field is given by the negative gradient of the Stark energy of the quantum level that they are in. This 
force can also be expressed as the product of the gradient of the electric field strength with an effective dipole moment, 
\textmu$_{\textrm{eff}}$. This effective dipole moment is the negative derivative of the Stark energy of the particular quantum state with 
respect to the strength of the electric field. In the presence of an electric field, every $J_{K_aK_c}$ level is split into $J+1$ 
different $M$-components, of which $M=0$ is non-degenerate whereas all the $M \neq 0$ components are doubly degenerate. 
The calculated effective dipole moments of the individual $M$ components of the three rotational levels mentioned above are 
shown as function of electric field strength in \autoref{fig:results:efields}. At low electric field strengths there are several avoided
crossings between levels of equal symmetry, resulting in a complicated dependence of \textmu$_{\textrm{eff}}$ on the strength of 
the electric field. At high electric field strengths, the effective dipole moments all converge to a value of 0.0759~cm$^{-1}$/(kV/cm), 
corresponding to the permanent electric dipole moment of benzonitrile of 4.52~Debye. In the selector, the molecules that are on 
stable trajectories experience electric fields in the range indicated by the shaded area in \autoref{fig:results:efields}. For benzonitrile 
molecules in the lowest rotational level, in particular, this implies that the value for the effective dipole moment is almost constant. It
is also seen that the effective dipole moment is the largest for the ground-state level, explaining why it is completely absent from the
transmission spectrum shown in the middle panel of \autoref{fig:results:bnspectraselection}. 

The effective dipole moment of the $J_{K_aK_c}M$ = $2_{02}2$ level is indistinguishable from that of the $0_{00}$ ground-state 
level in the electric fields inside the selector. The value of \textmu$_{\textrm{eff}}$ for the $M=1$ and $M=0$ components of the 
$2_{02}$ rotational level are lower, but are in turn almost identical to those of the $M=1$ and $M=0$ components of the $1_{01}$ 
rotational level, respectively. This grouping of effective dipole moments originates from the electric field dependence of the Stark 
shifts in the so-called pendular or harmonic librator limit~\cite{Loesch:JCP93:4779, Friedrich:ZPD18:153}; 
although the grouping of effective dipole moments  can be readily derived for a diatomic molecule~\cite{Bethlem:JPB39:R263} and even
for a symmetric-top molecule~\cite{Haertelt:JCP128:224313}, the correlation of the eigenenergies in the strong-field limit with those 
of the free rotor has not been rigorously analyzed for an asymmetric-top molecule like benzonitrile. 
Upon exiting the electric fields of the selector, the population in each of the $M$ components adiabatically 
transfers to the (degenerate) field-free rotational levels. In the LIF detection zone, therefore, the summed population of all $M$ 
components belonging to a certain rotational level is measured. This can nicely explain the degree of transmission of benzonitrile 
molecules in the $2_{02}$ level through the selector as observed in the lower spectrum in \autoref{fig:results:bnspectraselection}.
The integrated LIF signal intensity of ground-state molecules in this spectrum is seen to be decreased to about $3/4$ of its value in the
free-flight spectrum, whereas this is about $1/3$ for molecules in the $2_{02}$ level. Considering that 40\% of the molecules entering
the selector in the $2_{02}$ level will be in the $M=2$ component, the observed intensity in the spectrum is as expected when only 
the ground-state level and the $M=2$ component of the $2_{02}$ level contribute to the LIF signal behind the selector. From the weak
LIF signal of benzonitrile molecules in the $1_{01}$ level in this spectrum it can be deduced that molecules in the $M=1$ component
of this level (as well as of the $2_{02}$ level) have at least a factor five lower transmission than the ground-state level. This nicely 
demonstrates the degree of selectivity that can be obtained with the present selector as it is seen from \autoref{fig:results:efields} that 
the relative difference in the "electric-field-averaged" effective dipole moment of this $M=1$ component and the ground-state level is 
less than 10\%. 

\begin{figure}
   \centering
   \includegraphics[width=\linewidth]{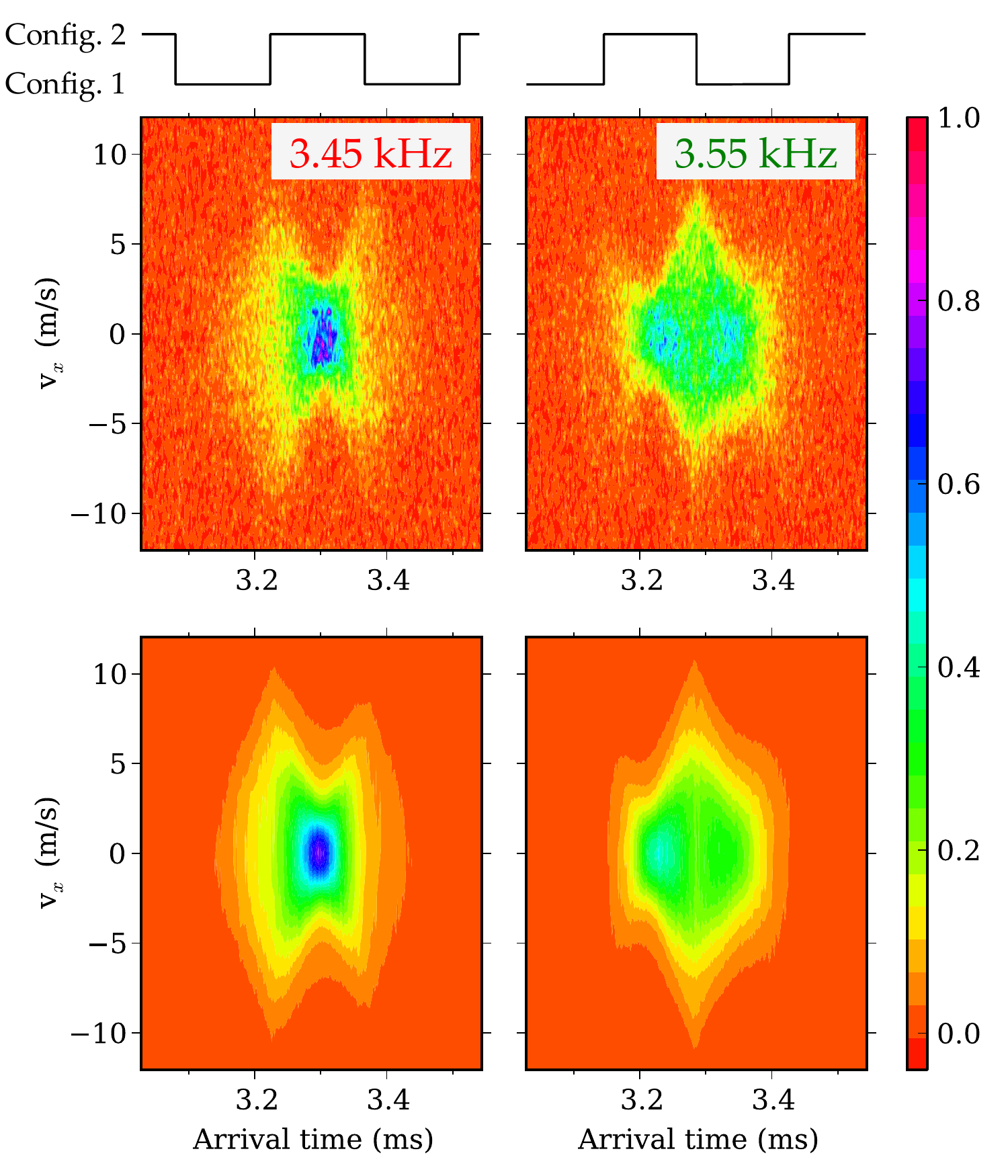}
   \caption{Upper row: Measured intensity of the LIF signal, plotted in a false-colour representation, as a function of the velocity in 
     the x-direction and as a function of the arrival time for two different ac switching frequencies. Data from \autoref{fig:results:linescans}.
     Lower row: Corresponding simulated arrival time distributions. The calculated velocity distributions are convoluted with a Lorentzian 
     profile with a FWHM width of 2.2~m/s.}
  \label{fig:results:linescans:theory}
\end{figure}

In \autoref{fig:results:linescans:theory} the experimental data shown in \autoref{fig:results:linescans} are compared to the outcome of 
trajectory calculations. In these calculations a pulse of molecules with an initial length of 4.0~cm (FWHM; Gaussian distribution),
with a mean forward velocity of 570~m/s and a velocity spread of 25~m/s (FWHM; Gaussian distribution) is assumed. The initial
transverse velocity distribution is described by a Gaussian distribution with a FWHM of 10~m/s, centered around zero velocity. The
dimensions of the molecular beam machine and the characteristics of the switching sequence as applied in the experiment are used as 
input. Effects due to the fringe fields near the ends of the electrodes of the selector are neglected in the simulations. The electric fields in the
selector are calculated using a finite element program package (COMSOL Multiphysics) and are assumed to be the same over the full 1640~mm
length of the selector. Switching between the two electric field configurations is assumed to be instantaneous. The voltages applied
to the electrodes have been set to $\pm$9.5~kV, just as used in the experiment. The electric field dependent dipole moment for the 
$0_{00}$ level as shown in \autoref{fig:results:efields} is taken, and the equations of motion are solved in sufficiently small time-steps. 
In the detection region, molecules in a 2.5~mm high interval are accepted, weighed by the duration that they are in the 2.5 mm diameter 
cylindrical interaction volume, i.e., assuming that the laser excitation step is not saturated, and the number of molecules are plotted as a function of 
their arrival time (binned in 2~\textmu s intervals) and as a function of their velocity component in the x-direction (binned in 0.3~m/s
intervals). To enable a direct comparison with the experimentally observed patterns, the calculated velocity distributions at a given 
arrival time are convoluted with a Lorentzian profile with a FWHM width of 2.2~m/s, thus taking the 8~MHz homogeneous broadening due to 
the finite lifetime of the electronically excited state, that is inherent in the measurements, into account. It is seen from
\autoref{fig:results:linescans:theory} that the experimentally observed patterns are quantitatively reproduced in the simulations, implying 
that in particular also the transverse velocity distributions of the molecules exiting the selector are described correctly.

\begin{figure}
   \centering
   \includegraphics[width=\linewidth]{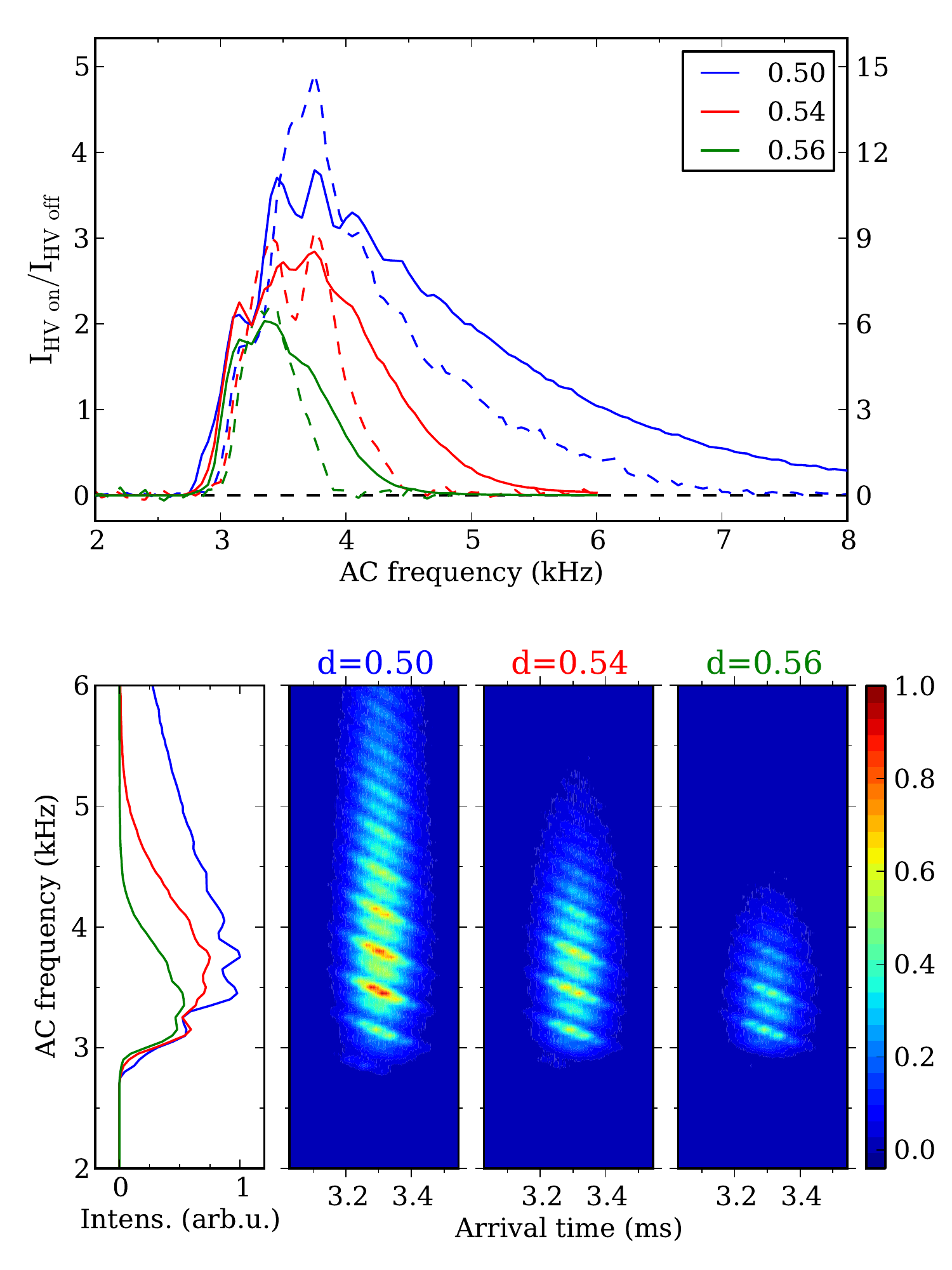}
   \caption{Upper row: Calculated time-integrated transmission curves for three different duty cycles (solid curves), compared to the 
     corresponding measured curves (dashed curves), already shown in \autoref{fig:results:efields}. The vertical scale for the measured 
     curves is on the left, while the one for the calculated ones is on the right.
     Lower row: Calculated number of ground-state benzonitrile molecules transmitted through the selector, plotted in a false-colour 
     representation, as a function of the ac switching frequency and as a function of arrival time for three different duty cycles. The panel 
     on the left shows the time-integrated signals as a function of ac switching frequency, also shown in the upper row. See text for further
     details.}
  \label{fig:results:frequscans}
\end{figure}

The calculated time-integrated and time-resolved number of ground-state benzonitrile molecules transmitted through the selector 
as a function of ac switching frequency is shown in \autoref{fig:results:frequscans}. Also in these calculations, the velocity 
distributions along the x-direction at a given arrival time are first convoluted with a Lorentzian profile with a FWHM width of 2.2~m/s 
and then only the molecules within a 0.5~m/s interval centered around 0~m/s are accepted, thereby mimicking the LIF detection of the
molecules in the actual experiment and enabling a direct comparison to the measurements shown in \autoref{fig:results:transmissioncurves}. 
In the comparison of the calculated and measured transmission curves shown in the upper row, it should be noted that the vertical scale 
of either set of curves, shown on the right and the left, respectively, is normalized to the situation in free flight. The observed overall 
transmission of the selector is seen to be about a factor of three lower than expected from the calculations. The shapes of the measured 
and the calculated curves are in good agreement, however. The observed modulations in the transmission curves and the low-frequency 
onset of the transmission of the selector are well reproduced in the calculations. At the high frequency end the experimentally observed 
transmission curves drop slightly faster than expected from the simulations. The observed change in the transmission characteristics with 
changing duty cycle is in quantitative agreement with the outcome of the trajectory calculations. The latter can therefore be used to 
quantify the \textmu/$\Delta$\textmu~resolution that can be obtained with the present selector.

\section{Conclusions}
\label{sec:conclusions}
In this work, we have demonstrated and characterized the performance of a state-of-the-art m/\textmu~selector for large molecules. 
Using a pulsed beam of benzonitrile molecules in combination with detection with a narrowband continuous wave laser, the 
quantum-state-specific transmission of the selector as well as the transverse velocity distribution of the molecules exiting the selector 
have been measured as a function of ac switching frequency. It has been demonstrated that by controlling the applied ac waveforms, 
a \textmu/$\Delta$\textmu~resolution can be obtained that is sufficient to exclusively transmit molecules in quantum states that have 
the same m/\textmu-value 
as the absolute ground state of benzonitrile. The measured transmission characteristics of the m/\textmu~selector are in quantitative 
agreement with the outcome of trajectory simulations; only at high frequencies the observed transmission is smaller than expected. 
It is well-known, however, that the transmission through the selector under these operating conditions is extremely sensitive to 
misalignments~\cite{Wohlfart:PRA77:031404,Bethlem:JPB39:R263}; fluctuations in the voltages applied to the electrodes might also 
play a role. Moreover, at these high 
frequencies the minute losses due to nonadiabatic transitions that might occur each time that the fields are being switched will be most 
notable. If we assume, for instance, that some 5\% of the ground-state molecules are lost to other quantum states each time that the 
fields are being switched while the molecules are inside the selector, we can bring both the observed overall transmission efficiency and 
the observed drop in transmission at high ac switching frequencies in perfect agreement with the calculations. 

The good agreement between the observed and calculated performance of the m/\textmu~selector is not only a testimony to the accurate 
mechanical design of the selector, but it also implies that trajectory calculations can be used to predict the outcome of future experiments. 
In the present selector, the number of full switching cycles that the molecules experience near optimum transmission is only about ten, which 
poses an intrinsic limitation to the attainable \textmu/$\Delta$\textmu~resolution. Nevertheless, trajectory calculations predict that for molecules 
with the mass of benzonitrile and with an effective dipole moment of around 4.5~Debye, the selector can be operated such that the 
overall transmission for levels with a constant effective dipole moment differing by 5\% can be made to differ by a factor of two; when
the \textmu/$\Delta$\textmu~resolution is defined in this way, a value of 20 is obtained. If a beam
containing mainly the species with the lower dipole moment of the two is desired, the selector can operate below the low-frequency cut-off for the
other species at a duty cycle of 0.50. A beam predominantly containing the species with the largest dipole moment of the two can only be
obtained at higher frequencies and at higher (or lower) duty cycles, thereby necessarily reducing the overall transmission somewhat. Apart from separating
quantum states of a certain molecule, the selector can also be used to separate molecules with different masses. It is expected, for instance,
that in the current setup benzonitrile-Ar clusters will be present in the molecular beam as well. These clusters are expected to have a dipole 
moment that is very similar to that of the bare benzonitrile molecule, but their rotational level structure will be more dense and their mass is 
considerably higher. With the selector operating at a frequency below the low-frequency cut-off for benzonitrile, with a duty cycle of 0.50, it 
should be possible, therefore, to selectively transmit these higher mass clusters. A further possibility to increase the \textmu/$\Delta$\textmu~resolution
might be the modulation of the high voltages applied to the electrodes with specific frequencies to selectively eject molecules with unwanted
m/\textmu~ratios via parametric amplification of their transverse motion. Studies are underway to investigate these applications of the 
m/\textmu~selector.

The m/\textmu~selector provides samples of highly oriented molecules. To maintain a defined orientation for a subsequent experiment, a static
electric guiding field of about 25~kV/cm is needed at the exit of the selector, which is well feasible.
It should be mentioned that as an alternative to the m/\textmu~selector, a static "two-wire" deflector has been used to prepare pure 
samples of large molecules for orientation and alignment studies \cite{Holmegaard:PRL102:023001,Filsinger:JCP131:064309}
as well as to separate structural isomers of molecules~\cite{Filsinger:AC121:7033}. Although the implementation of the deflector is more 
straightforward than that of the m/\textmu~selector, its application is also more limited; the difference in applicability between the two devices 
is similar to the difference in applicability between a static ion deflector (a separator or bender) and a quadrupole mass-filter.

\section{Acknowledgments}
\label{sec:acknowledgments}
The design of the electronics by G. Heyne, V. Platschkowski and P. Schlecht has been crucial for this work. We acknowledge the software 
support by U. Hoppe and H. Junkes.

\bibliography{string,mp}

\begin{thebibliography}{28}%
\makeatletter
\providecommand \@ifxundefined [1]{%
 \@ifx{#1\undefined}
}%
\providecommand \@ifnum [1]{%
 \ifnum #1\expandafter \@firstoftwo
 \else \expandafter \@secondoftwo
 \fi
}%
\providecommand \@ifx [1]{%
 \ifx #1\expandafter \@firstoftwo
 \else \expandafter \@secondoftwo
 \fi
}%
\providecommand \natexlab [1]{#1}%
\providecommand \enquote  [1]{``#1''}%
\providecommand \bibnamefont  [1]{#1}%
\providecommand \bibfnamefont [1]{#1}%
\providecommand \citenamefont [1]{#1}%
\providecommand \href@noop [0]{\@secondoftwo}%
\providecommand \href [0]{\begingroup \@sanitize@url \@href}%
\providecommand \@href[1]{\@@startlink{#1}\@@href}%
\providecommand \@@href[1]{\endgroup#1\@@endlink}%
\providecommand \@sanitize@url [0]{\catcode `\\12\catcode `\$12\catcode
  `\&12\catcode `\#12\catcode `\^12\catcode `\_12\catcode `\%12\relax}%
\providecommand \@@startlink[1]{}%
\providecommand \@@endlink[0]{}%
\providecommand \url  [0]{\begingroup\@sanitize@url \@url }%
\providecommand \@url [1]{\endgroup\@href {#1}{\urlprefix }}%
\providecommand \urlprefix  [0]{URL }%
\providecommand \Eprint [0]{\href }%
\@ifxundefined \urlstyle {%
  \providecommand \doi  [0]{\begingroup \@sanitize@url \@doi}%
  \providecommand \@doi [1]{\endgroup \@@startlink {\doibase
  #1}doi:\discretionary {}{}{}#1\@@endlink }%
}{%
  \providecommand \doi  [0]{doi:\discretionary{}{}{}\begingroup
  \urlstyle{rm}\Url }%
}%
\providecommand \doibase [0]{http://dx.doi.org/}%
\providecommand \Doi [0]{\begingroup \@sanitize@url \@Doi }%
\providecommand \@Doi  [1]{\endgroup\@@startlink{\doibase#1}\@@Doi}%
\providecommand \@@Doi [1]{#1\@@endlink}%
\providecommand \selectlanguage [0]{\@gobble}%
\providecommand \bibinfo  [0]{\@secondoftwo}%
\providecommand \bibfield  [0]{\@secondoftwo}%
\providecommand \translation [1]{[#1]}%
\providecommand \BibitemOpen [0]{}%
\providecommand \bibitemStop [0]{}%
\providecommand \bibitemNoStop [0]{.\EOS\space}%
\providecommand \EOS [0]{\spacefactor3000\relax}%
\providecommand \BibitemShut  [1]{\csname bibitem#1\endcsname}%
\bibitem [{\citenamefont {Scoles}(1992)}]{Scoles:MolBeam:1and2}%
  \BibitemOpen
  \bibinfo {editor} {\bibfnamefont {G.}~\bibnamefont {Scoles}},\ ed.,\
  \href@noop {} {\emph {\bibinfo {title} {Atomic and molecular beam
  methods}}},\ Vol.\ \bibinfo {volume} {1 \& 2}\ (\bibinfo  {publisher} {Oxford
  University Press},\ \bibinfo {address} {New York, NY, USA},\ \bibinfo {year}
  {1988 \& 1992})\ ISBN \bibinfo {isbn} {0195042808}\BibitemShut {NoStop}%
\bibitem [{\citenamefont {Levy}(1981)}]{Levy:Science214:263}%
  \BibitemOpen
  \bibfield  {author} {\bibinfo {author} {\bibfnamefont {D.~H.}\ \bibnamefont
  {Levy}},\ }\Doi {10.1126/science.214.4518.263} {\bibfield  {journal}
  {\bibinfo  {journal} {Science},\ }\textbf {\bibinfo {volume} {214}},\
  \bibinfo {pages} {263} (\bibinfo {year} {1981})}\BibitemShut {NoStop}%
\bibitem [{\citenamefont {Rizzo}\ \emph {et~al.}(1985)\citenamefont {Rizzo},
  \citenamefont {Park}, \citenamefont {Peteanu},\ and\ \citenamefont
  {Levy}}]{Rizzo:JCP83:4819}%
  \BibitemOpen
  \bibfield  {author} {\bibinfo {author} {\bibfnamefont {T.~R.}\ \bibnamefont
  {Rizzo}}, \bibinfo {author} {\bibfnamefont {Y.~D.}\ \bibnamefont {Park}},
  \bibinfo {author} {\bibfnamefont {L.}~\bibnamefont {Peteanu}}, \ and\
  \bibinfo {author} {\bibfnamefont {D.~H.}\ \bibnamefont {Levy}},\ }\Doi
  {10.1063/1.449009} {\bibfield  {journal} {\bibinfo  {journal} {J.\ Chem.\
  Phys.},\ }\textbf {\bibinfo {volume} {83}},\ \bibinfo {pages} {4819}
  (\bibinfo {year} {1985})}\BibitemShut {NoStop}%
\bibitem [{\citenamefont {van~de Meerakker}\ \emph {et~al.}(2008)\citenamefont
  {van~de Meerakker}, \citenamefont {Bethlem},\ and\ \citenamefont
  {Meijer}}]{Meerakker:NatPhys4:595}%
  \BibitemOpen
  \bibfield  {author} {\bibinfo {author} {\bibfnamefont {S.~Y.~T.}\
  \bibnamefont {van~de Meerakker}}, \bibinfo {author} {\bibfnamefont {H.~L.}\
  \bibnamefont {Bethlem}}, \ and\ \bibinfo {author} {\bibfnamefont
  {G.}~\bibnamefont {Meijer}},\ }\Doi {10.1038/nphys1031} {\bibfield  {journal}
  {\bibinfo  {journal} {Nature Phys.},\ }\textbf {\bibinfo {volume} {4}},\
  \bibinfo {pages} {595} (\bibinfo {year} {2008})}\BibitemShut {NoStop}%
\bibitem [{\citenamefont {Gordon}\ \emph {et~al.}(1955)\citenamefont {Gordon},
  \citenamefont {Zeiger},\ and\ \citenamefont {Townes}}]{Gordon:PR99:1264}%
  \BibitemOpen
  \bibfield  {author} {\bibinfo {author} {\bibfnamefont {J.~P.}\ \bibnamefont
  {Gordon}}, \bibinfo {author} {\bibfnamefont {H.~J.}\ \bibnamefont {Zeiger}},
  \ and\ \bibinfo {author} {\bibfnamefont {C.~H.}\ \bibnamefont {Townes}},\
  }\Doi {10.1103/PhysRev.99.1264} {\bibfield  {journal} {\bibinfo  {journal}
  {Phys.\ Rev.},\ }\textbf {\bibinfo {volume} {99}},\ \bibinfo {pages} {1264}
  (\bibinfo {year} {1955})}\BibitemShut {NoStop}%
\bibitem [{\citenamefont {Auerbach}\ \emph {et~al.}(1966)\citenamefont
  {Auerbach}, \citenamefont {Bromberg},\ and\ \citenamefont
  {Wharton}}]{Auerbach:JCP45:2160}%
  \BibitemOpen
  \bibfield  {author} {\bibinfo {author} {\bibfnamefont {D.}~\bibnamefont
  {Auerbach}}, \bibinfo {author} {\bibfnamefont {E.~E.~A.}\ \bibnamefont
  {Bromberg}}, \ and\ \bibinfo {author} {\bibfnamefont {L.}~\bibnamefont
  {Wharton}},\ }\Doi {10.1063/1.1727902} {\bibfield  {journal} {\bibinfo
  {journal} {J.\ Chem.\ Phys.},\ }\textbf {\bibinfo {volume} {45}},\ \bibinfo
  {pages} {2160} (\bibinfo {year} {1966})}\BibitemShut {NoStop}%
\bibitem [{\citenamefont {Courant}\ and\ \citenamefont
  {Snyder}(1958)}]{Courant:AnnPhys3:1}%
  \BibitemOpen
  \bibfield  {author} {\bibinfo {author} {\bibfnamefont {E.~D.}\ \bibnamefont
  {Courant}}\ and\ \bibinfo {author} {\bibfnamefont {H.~S.}\ \bibnamefont
  {Snyder}},\ }\Doi {10.1016/0003-4916(58)90012-5} {\bibfield  {journal}
  {\bibinfo  {journal} {Ann. Phys.},\ }\textbf {\bibinfo {volume} {3}},\
  \bibinfo {pages} {1} (\bibinfo {year} {1958})}\BibitemShut {NoStop}%
\bibitem [{\citenamefont {Kakati}\ and\ \citenamefont
  {Lain\'e}(1967)}]{Kakati:PLA24:676}%
  \BibitemOpen
  \bibfield  {author} {\bibinfo {author} {\bibfnamefont {D.}~\bibnamefont
  {Kakati}}\ and\ \bibinfo {author} {\bibfnamefont {D.~C.}\ \bibnamefont
  {Lain\'e}},\ }\Doi {10.1016/0375-9601(67)91022-5} {\bibfield  {journal}
  {\bibinfo  {journal} {Phys.\ Lett.\ A},\ }\textbf {\bibinfo {volume} {24}},\
  \bibinfo {pages} {676} (\bibinfo {year} {1967})}\BibitemShut {NoStop}%
\bibitem [{\citenamefont {Bethlem}\ \emph {et~al.}(2002)\citenamefont
  {Bethlem}, \citenamefont {van Roij}, \citenamefont {Jongma},\ and\
  \citenamefont {Meijer}}]{Bethlem:PRL88:133003}%
  \BibitemOpen
  \bibfield  {author} {\bibinfo {author} {\bibfnamefont {H.~L.}\ \bibnamefont
  {Bethlem}}, \bibinfo {author} {\bibfnamefont {A.~J.~A.}\ \bibnamefont {van
  Roij}}, \bibinfo {author} {\bibfnamefont {R.~T.}\ \bibnamefont {Jongma}}, \
  and\ \bibinfo {author} {\bibfnamefont {G.}~\bibnamefont {Meijer}},\ }\Doi
  {10.1103/PhysRevLett.88.133003} {\bibfield  {journal} {\bibinfo  {journal}
  {Phys.\ Rev.\ Lett.},\ }\textbf {\bibinfo {volume} {88}},\ \bibinfo {pages}
  {133003} (\bibinfo {year} {2002})}\BibitemShut {NoStop}%
\bibitem [{\citenamefont {van Veldhoven}\ \emph {et~al.}(2005)\citenamefont
  {van Veldhoven}, \citenamefont {Bethlem},\ and\ \citenamefont
  {Meijer}}]{Veldhoven:PRL94:083001}%
  \BibitemOpen
  \bibfield  {author} {\bibinfo {author} {\bibfnamefont {J.}~\bibnamefont {van
  Veldhoven}}, \bibinfo {author} {\bibfnamefont {H.~L.}\ \bibnamefont
  {Bethlem}}, \ and\ \bibinfo {author} {\bibfnamefont {G.}~\bibnamefont
  {Meijer}},\ }\Doi {10.1103/PhysRevLett.94.083001} {\bibfield  {journal}
  {\bibinfo  {journal} {Phys.\ Rev.\ Lett.},\ }\textbf {\bibinfo {volume}
  {94}},\ \bibinfo {pages} {083001} (\bibinfo {year} {2005})}\BibitemShut
  {NoStop}%
\bibitem [{\citenamefont {Junglen}\ \emph {et~al.}(2004)\citenamefont
  {Junglen}, \citenamefont {Rieger}, \citenamefont {Rangwala}, \citenamefont
  {Pinkse},\ and\ \citenamefont {Rempe}}]{Junglen:PRL92:223001}%
  \BibitemOpen
  \bibfield  {author} {\bibinfo {author} {\bibfnamefont {T.}~\bibnamefont
  {Junglen}}, \bibinfo {author} {\bibfnamefont {T.}~\bibnamefont {Rieger}},
  \bibinfo {author} {\bibfnamefont {S.~A.}\ \bibnamefont {Rangwala}}, \bibinfo
  {author} {\bibfnamefont {P.~W.~H.}\ \bibnamefont {Pinkse}}, \ and\ \bibinfo
  {author} {\bibfnamefont {G.}~\bibnamefont {Rempe}},\ }\Doi
  {10.1103/PhysRevLett.92.223001} {\bibfield  {journal} {\bibinfo  {journal}
  {Phys.\ Rev.\ Lett.},\ }\textbf {\bibinfo {volume} {92}},\ \bibinfo {pages}
  {223001} (\bibinfo {year} {2004})}\BibitemShut {NoStop}%
\bibitem [{\citenamefont {Wall}\ \emph {et~al.}(2009)\citenamefont {Wall},
  \citenamefont {Armitage}, \citenamefont {Hudson}, \citenamefont {Sauer},
  \citenamefont {Dyne}, \citenamefont {Hinds},\ and\ \citenamefont
  {Tarbutt}}]{Wall:PRA80:043407}%
  \BibitemOpen
  \bibfield  {author} {\bibinfo {author} {\bibfnamefont {T.~E.}\ \bibnamefont
  {Wall}}, \bibinfo {author} {\bibfnamefont {S.}~\bibnamefont {Armitage}},
  \bibinfo {author} {\bibfnamefont {J.~J.}\ \bibnamefont {Hudson}}, \bibinfo
  {author} {\bibfnamefont {B.~E.}\ \bibnamefont {Sauer}}, \bibinfo {author}
  {\bibfnamefont {J.~M.}\ \bibnamefont {Dyne}}, \bibinfo {author}
  {\bibfnamefont {E.~A.}\ \bibnamefont {Hinds}}, \ and\ \bibinfo {author}
  {\bibfnamefont {M.~R.}\ \bibnamefont {Tarbutt}},\ }\Doi
  {10.1103/PhysRevA.80.043407} {\bibfield  {journal} {\bibinfo  {journal}
  {Phys.\ Rev.\ A},\ }\textbf {\bibinfo {volume} {80}},\ \bibinfo {pages}
  {043407} (\bibinfo {year} {2009})}\BibitemShut {NoStop}%
\bibitem [{\citenamefont {Wohlfart}\ \emph
  {et~al.}(2008){\natexlab{a}}\citenamefont {Wohlfart}, \citenamefont
  {Gr\"atz}, \citenamefont {Filsinger}, \citenamefont {Haak}, \citenamefont
  {Meijer},\ and\ \citenamefont {K\"upper}}]{Wohlfart:PRA77:031404}%
  \BibitemOpen
  \bibfield  {author} {\bibinfo {author} {\bibfnamefont {K.}~\bibnamefont
  {Wohlfart}}, \bibinfo {author} {\bibfnamefont {F.}~\bibnamefont {Gr\"atz}},
  \bibinfo {author} {\bibfnamefont {F.}~\bibnamefont {Filsinger}}, \bibinfo
  {author} {\bibfnamefont {H.}~\bibnamefont {Haak}}, \bibinfo {author}
  {\bibfnamefont {G.}~\bibnamefont {Meijer}}, \ and\ \bibinfo {author}
  {\bibfnamefont {J.}~\bibnamefont {K\"upper}},\ }\Doi
  {10.1103/PhysRevA.77.031404} {\bibfield  {journal} {\bibinfo  {journal}
  {Phys.\ Rev.\ A},\ }\textbf {\bibinfo {volume} {77}},\ \bibinfo {pages}
  {031404(R)} (\bibinfo {year} {2008}{\natexlab{a}})}\BibitemShut {NoStop}%
\bibitem [{\citenamefont {Filsinger}\ \emph {et~al.}(2008)\citenamefont
  {Filsinger}, \citenamefont {Erlekam}, \citenamefont {von Helden},
  \citenamefont {K\"upper},\ and\ \citenamefont
  {Meijer}}]{Filsinger:PRL100:133003}%
  \BibitemOpen
  \bibfield  {author} {\bibinfo {author} {\bibfnamefont {F.}~\bibnamefont
  {Filsinger}}, \bibinfo {author} {\bibfnamefont {U.}~\bibnamefont {Erlekam}},
  \bibinfo {author} {\bibfnamefont {G.}~\bibnamefont {von Helden}}, \bibinfo
  {author} {\bibfnamefont {J.}~\bibnamefont {K\"upper}}, \ and\ \bibinfo
  {author} {\bibfnamefont {G.}~\bibnamefont {Meijer}},\ }\Doi
  {10.1103/PhysRevLett.100.133003} {\bibfield  {journal} {\bibinfo  {journal}
  {Phys.\ Rev.\ Lett.},\ }\textbf {\bibinfo {volume} {100}},\ \bibinfo {pages}
  {133003} (\bibinfo {year} {2008})}\BibitemShut {NoStop}%
\bibitem [{\citenamefont {Filsinger}\ \emph {et~al.}(2010)\citenamefont
  {Filsinger}, \citenamefont {Putzke}, \citenamefont {Haak}, \citenamefont
  {Meijer},\ and\ \citenamefont {K\"upper}}]{PhysRevA.82.052513}%
  \BibitemOpen
  \bibfield  {author} {\bibinfo {author} {\bibfnamefont {F.}~\bibnamefont
  {Filsinger}}, \bibinfo {author} {\bibfnamefont {S.}~\bibnamefont {Putzke}},
  \bibinfo {author} {\bibfnamefont {H.}~\bibnamefont {Haak}}, \bibinfo {author}
  {\bibfnamefont {G.}~\bibnamefont {Meijer}}, \ and\ \bibinfo {author}
  {\bibfnamefont {J.}~\bibnamefont {K\"upper}},\ }\Doi
  {10.1103/PhysRevA.82.052513} {\bibfield  {journal} {\bibinfo  {journal}
  {Phys. Rev. A},\ }\textbf {\bibinfo {volume} {82}},\ \bibinfo {pages}
  {052513} (\bibinfo {year} {2010})}\BibitemShut {NoStop}%
\bibitem [{\citenamefont {Wohlfart}\ \emph
  {et~al.}(2008){\natexlab{b}}\citenamefont {Wohlfart}, \citenamefont
  {Schnell}, \citenamefont {Grabow},\ and\ \citenamefont
  {K\"upper}}]{Wohlfart:JMolSpec247:119}%
  \BibitemOpen
  \bibfield  {author} {\bibinfo {author} {\bibfnamefont {K.}~\bibnamefont
  {Wohlfart}}, \bibinfo {author} {\bibfnamefont {M.}~\bibnamefont {Schnell}},
  \bibinfo {author} {\bibfnamefont {J.-U.}\ \bibnamefont {Grabow}}, \ and\
  \bibinfo {author} {\bibfnamefont {J.}~\bibnamefont {K\"upper}},\ }\Doi
  {10.1016/j.jms.2007.10.006} {\bibfield  {journal} {\bibinfo  {journal} {J.\
  Mol.\ Spec.},\ }\textbf {\bibinfo {volume} {247}},\ \bibinfo {pages} {119}
  (\bibinfo {year} {2008}{\natexlab{b}})}\BibitemShut {NoStop}%
\bibitem [{\citenamefont {Borst}\ \emph {et~al.}(2001)\citenamefont {Borst},
  \citenamefont {Korter},\ and\ \citenamefont {Pratt}}]{Borst:CPL350:485}%
  \BibitemOpen
  \bibfield  {author} {\bibinfo {author} {\bibfnamefont {D.~R.}\ \bibnamefont
  {Borst}}, \bibinfo {author} {\bibfnamefont {T.~M.}\ \bibnamefont {Korter}}, \
  and\ \bibinfo {author} {\bibfnamefont {D.~W.}\ \bibnamefont {Pratt}},\ }\Doi
  {10.1016/S0009-2614(01)01344-6} {\bibfield  {journal} {\bibinfo  {journal}
  {Chem.\ Phys.\ Lett.},\ }\textbf {\bibinfo {volume} {350}},\ \bibinfo {pages}
  {485} (\bibinfo {year} {2001})}\BibitemShut {NoStop}%
\bibitem [{\citenamefont {Western}()}]{Western:pgopher}%
  \BibitemOpen
  \bibfield  {author} {\bibinfo {author} {\bibfnamefont {C.~M.}\ \bibnamefont
  {Western}},\ }\href@noop {} {\enquote {\bibinfo {title} {{PGOPHER}, a program
  for simulating rotational structure},}\ }\bibinfo {note} {{U}niversity of
  {B}ristol, {B}ristol, {UK},
  URL:~\url{http://pgopher.chm.bris.ac.uk}}\BibitemShut {NoStop}%
\bibitem [{\citenamefont {Kobayashi}\ \emph {et~al.}(1987)\citenamefont
  {Kobayashi}, \citenamefont {Honma}, \citenamefont {Kajimoto},\ and\
  \citenamefont {Tsuchiya}}]{Kobayashi:JCP86:1111}%
  \BibitemOpen
  \bibfield  {author} {\bibinfo {author} {\bibfnamefont {T.}~\bibnamefont
  {Kobayashi}}, \bibinfo {author} {\bibfnamefont {K.}~\bibnamefont {Honma}},
  \bibinfo {author} {\bibfnamefont {O.}~\bibnamefont {Kajimoto}}, \ and\
  \bibinfo {author} {\bibfnamefont {S.}~\bibnamefont {Tsuchiya}},\ }\href
  {http://link.aip.org/link/?JCP/86/1111/1} {\bibfield  {journal} {\bibinfo
  {journal} {J.\ Chem.\ Phys.},\ }\textbf {\bibinfo {volume} {86}},\ \bibinfo
  {pages} {1111} (\bibinfo {year} {1987})}\BibitemShut {NoStop}%
\bibitem [{\citenamefont {Lahmani}\ \emph {et~al.}(1989)\citenamefont
  {Lahmani}, \citenamefont {Lardeux-Dedonder},\ and\ \citenamefont
  {Zehnacker}}]{Lahmani:LaserChem10:41}%
  \BibitemOpen
  \bibfield  {author} {\bibinfo {author} {\bibfnamefont {F.}~\bibnamefont
  {Lahmani}}, \bibinfo {author} {\bibfnamefont {C.}~\bibnamefont
  {Lardeux-Dedonder}}, \ and\ \bibinfo {author} {\bibfnamefont
  {A.}~\bibnamefont {Zehnacker}},\ }\Doi {10.1155/1989/65147} {\bibfield
  {journal} {\bibinfo  {journal} {Laser Chemistry},\ }\textbf {\bibinfo
  {volume} {10}},\ \bibinfo {pages} {41} (\bibinfo {year} {1989})}\BibitemShut
  {NoStop}%
\bibitem [{\citenamefont {Bethlem}\ \emph {et~al.}(2006)\citenamefont
  {Bethlem}, \citenamefont {Tarbutt}, \citenamefont {K\"upper}, \citenamefont
  {Carty}, \citenamefont {Wohlfart}, \citenamefont {Hinds},\ and\ \citenamefont
  {Meijer}}]{Bethlem:JPB39:R263}%
  \BibitemOpen
  \bibfield  {author} {\bibinfo {author} {\bibfnamefont {H.~L.}\ \bibnamefont
  {Bethlem}}, \bibinfo {author} {\bibfnamefont {M.~R.}\ \bibnamefont
  {Tarbutt}}, \bibinfo {author} {\bibfnamefont {J.}~\bibnamefont {K\"upper}},
  \bibinfo {author} {\bibfnamefont {D.}~\bibnamefont {Carty}}, \bibinfo
  {author} {\bibfnamefont {K.}~\bibnamefont {Wohlfart}}, \bibinfo {author}
  {\bibfnamefont {E.~A.}\ \bibnamefont {Hinds}}, \ and\ \bibinfo {author}
  {\bibfnamefont {G.}~\bibnamefont {Meijer}},\ }\Doi
  {10.1088/0953-4075/39/16/R01} {\bibfield  {journal} {\bibinfo  {journal} {J.\
  Phys.\ B},\ }\textbf {\bibinfo {volume} {39}},\ \bibinfo {pages} {R263}
  (\bibinfo {year} {2006})}\BibitemShut {NoStop}%
\bibitem [{\citenamefont {Schlunk}\ \emph {et~al.}(2007)\citenamefont
  {Schlunk}, \citenamefont {Marian}, \citenamefont {Geng}, \citenamefont
  {Mosk}, \citenamefont {Meijer},\ and\ \citenamefont
  {Sch\"ollkopf}}]{Schlunk:PRL98:223002}%
  \BibitemOpen
  \bibfield  {author} {\bibinfo {author} {\bibfnamefont {S.}~\bibnamefont
  {Schlunk}}, \bibinfo {author} {\bibfnamefont {A.}~\bibnamefont {Marian}},
  \bibinfo {author} {\bibfnamefont {P.}~\bibnamefont {Geng}}, \bibinfo {author}
  {\bibfnamefont {A.~P.}\ \bibnamefont {Mosk}}, \bibinfo {author}
  {\bibfnamefont {G.}~\bibnamefont {Meijer}}, \ and\ \bibinfo {author}
  {\bibfnamefont {W.}~\bibnamefont {Sch\"ollkopf}},\ }\Doi
  {10.1103/PhysRevLett.98.223002} {\bibfield  {journal} {\bibinfo  {journal}
  {Phys.\ Rev.\ Lett.},\ }\textbf {\bibinfo {volume} {98}} (\bibinfo {year}
  {2007})},\ \doi {10.1103/PhysRevLett.98.223002}\BibitemShut {NoStop}%
\bibitem [{\citenamefont {Loesch}\ and\ \citenamefont
  {Remscheid}(1990)}]{Loesch:JCP93:4779}%
  \BibitemOpen
  \bibfield  {author} {\bibinfo {author} {\bibfnamefont {H.~J.}\ \bibnamefont
  {Loesch}}\ and\ \bibinfo {author} {\bibfnamefont {A.}~\bibnamefont
  {Remscheid}},\ }\Doi {10.1063/1.458668} {\bibfield  {journal} {\bibinfo
  {journal} {J.\ Chem.\ Phys.},\ }\textbf {\bibinfo {volume} {93}},\ \bibinfo
  {pages} {4779} (\bibinfo {year} {1990})}\BibitemShut {NoStop}%
\bibitem [{\citenamefont {Friedrich}\ and\ \citenamefont
  {Herschbach}(1991)}]{Friedrich:ZPD18:153}%
  \BibitemOpen
  \bibfield  {author} {\bibinfo {author} {\bibfnamefont {B.}~\bibnamefont
  {Friedrich}}\ and\ \bibinfo {author} {\bibfnamefont {D.~R.}\ \bibnamefont
  {Herschbach}},\ }\Doi {10.1007/BF01437441} {\bibfield  {journal} {\bibinfo
  {journal} {Z.\ Phys.\ D},\ }\textbf {\bibinfo {volume} {18}},\ \bibinfo
  {pages} {153} (\bibinfo {year} {1991})}\BibitemShut {NoStop}%
\bibitem [{\citenamefont {Hartelt}\ and\ \citenamefont
  {Friedrich}(2008)}]{Haertelt:JCP128:224313}%
  \BibitemOpen
  \bibfield  {author} {\bibinfo {author} {\bibfnamefont {M.}~\bibnamefont
  {Hartelt}}\ and\ \bibinfo {author} {\bibfnamefont {B.}~\bibnamefont
  {Friedrich}},\ }\Doi {10.1063/1.2929850} {\bibfield  {journal} {\bibinfo
  {journal} {The Journal of Chemical Physics},\ }\textbf {\bibinfo {volume}
  {128}},\ \bibinfo {eid} {224313} (\bibinfo {year} {2008})}\BibitemShut
  {NoStop}%
\bibitem [{\citenamefont {Holmegaard}\ \emph {et~al.}(2009)\citenamefont
  {Holmegaard}, \citenamefont {Nielsen}, \citenamefont {Nevo}, \citenamefont
  {Stapelfeldt}, \citenamefont {Filsinger}, \citenamefont {K\"upper},\ and\
  \citenamefont {Meijer}}]{Holmegaard:PRL102:023001}%
  \BibitemOpen
  \bibfield  {author} {\bibinfo {author} {\bibfnamefont {L.}~\bibnamefont
  {Holmegaard}}, \bibinfo {author} {\bibfnamefont {J.~H.}\ \bibnamefont
  {Nielsen}}, \bibinfo {author} {\bibfnamefont {I.}~\bibnamefont {Nevo}},
  \bibinfo {author} {\bibfnamefont {H.}~\bibnamefont {Stapelfeldt}}, \bibinfo
  {author} {\bibfnamefont {F.}~\bibnamefont {Filsinger}}, \bibinfo {author}
  {\bibfnamefont {J.}~\bibnamefont {K\"upper}}, \ and\ \bibinfo {author}
  {\bibfnamefont {G.}~\bibnamefont {Meijer}},\ }\Doi
  {10.1103/PhysRevLett.102.023001} {\bibfield  {journal} {\bibinfo  {journal}
  {Phys.\ Rev.\ Lett.},\ }\textbf {\bibinfo {volume} {102}},\ \bibinfo {pages}
  {023001} (\bibinfo {year} {2009})}\BibitemShut {NoStop}%
\bibitem [{\citenamefont {Filsinger}\ \emph
  {et~al.}(2009){\natexlab{a}}\citenamefont {Filsinger}, \citenamefont
  {K\"upper}, \citenamefont {Meijer}, \citenamefont {Holmegaard}, \citenamefont
  {Nielsen}, \citenamefont {Nevo}, \citenamefont {Hansen},\ and\ \citenamefont
  {Stapelfeldt}}]{Filsinger:JCP131:064309}%
  \BibitemOpen
  \bibfield  {author} {\bibinfo {author} {\bibfnamefont {F.}~\bibnamefont
  {Filsinger}}, \bibinfo {author} {\bibfnamefont {J.}~\bibnamefont {K\"upper}},
  \bibinfo {author} {\bibfnamefont {G.}~\bibnamefont {Meijer}}, \bibinfo
  {author} {\bibfnamefont {L.}~\bibnamefont {Holmegaard}}, \bibinfo {author}
  {\bibfnamefont {J.~H.}\ \bibnamefont {Nielsen}}, \bibinfo {author}
  {\bibfnamefont {I.}~\bibnamefont {Nevo}}, \bibinfo {author} {\bibfnamefont
  {J.~L.}\ \bibnamefont {Hansen}}, \ and\ \bibinfo {author} {\bibfnamefont
  {H.}~\bibnamefont {Stapelfeldt}},\ }\Doi {10.1063/1.3194287} {\bibfield
  {journal} {\bibinfo  {journal} {J.\ Chem.\ Phys.},\ }\textbf {\bibinfo
  {volume} {131}},\ \bibinfo {pages} {064309} (\bibinfo {year}
  {2009}{\natexlab{a}})}\BibitemShut {NoStop}%
\bibitem [{\citenamefont {Filsinger}\ \emph
  {et~al.}(2009){\natexlab{b}}\citenamefont {Filsinger}, \citenamefont
  {K\"upper}, \citenamefont {Meijer}, \citenamefont {Hansen}, \citenamefont
  {Maurer}, \citenamefont {Nielsen}, \citenamefont {Holmegaard},\ and\
  \citenamefont {Stapelfeldt}}]{Filsinger:AC121:7033}%
  \BibitemOpen
  \bibfield  {author} {\bibinfo {author} {\bibfnamefont {F.}~\bibnamefont
  {Filsinger}}, \bibinfo {author} {\bibfnamefont {J.}~\bibnamefont {K\"upper}},
  \bibinfo {author} {\bibfnamefont {G.}~\bibnamefont {Meijer}}, \bibinfo
  {author} {\bibfnamefont {J.~L.}\ \bibnamefont {Hansen}}, \bibinfo {author}
  {\bibfnamefont {J.}~\bibnamefont {Maurer}}, \bibinfo {author} {\bibfnamefont
  {J.~H.}\ \bibnamefont {Nielsen}}, \bibinfo {author} {\bibfnamefont
  {L.}~\bibnamefont {Holmegaard}}, \ and\ \bibinfo {author} {\bibfnamefont
  {H.}~\bibnamefont {Stapelfeldt}},\ }\Doi {10.1002/ange.200902650} {\bibfield
  {journal} {\bibinfo  {journal} {Angew.\ Chem.},\ }\textbf {\bibinfo {volume}
  {121}},\ \bibinfo {pages} {7033} (\bibinfo {year}
  {2009}{\natexlab{b}})}\BibitemShut {NoStop}%
\end{thebibliography}%

\end{document}